\documentclass[11pt,a4paper]{article}
\pdfoutput=1

\usepackage{jcappub}
\usepackage{booktabs}
\usepackage{multirow}
\usepackage{bm}
\usepackage{cleveref}
\usepackage{xcolor}
\usepackage{appendix}

\begin{document}

\title{Field-Level Baryon Acoustic Oscillation Reconstruction of the DESI DR1 Luminous Red Galaxies with Linear Field Transformer (LiFT)}

\author[a,b]{Liam Parker,}
\author[a,b]{Uroš Seljak,}
\author[c,d,e,f]{Adrian E. Bayer,}
\author[a,b]{Richard Feder,}
\author[a,b]{and David Valcin}

\affiliation[a]{Department of Physics, University of California, Berkeley, \\
366 LeConte Hall MC 7300, Berkeley, CA 94720-7300, U.S.A}
\affiliation[b]{Physics Division, Lawrence Berkeley National Laboratory, \\
1 Cyclotron Road, Berkeley, CA 94720, USA}
\affiliation[c]{The NSF AI Institute for Artificial Intelligence and Fundamental Interactions,\\ 
Cambridge, MA 02139, USA}
\affiliation[d]{Laboratory for Nuclear Science, Massachusetts Institute of Technology,\\ 
Cambridge, MA 02139, USA}
\affiliation[e]{Center for Astrophysics $|$ Harvard \& Smithsonian,
60 Garden Street,\\
Cambridge, MA 02138, USA}
\affiliation[f]{Perimeter Institute for Theoretical Physics,\\
31 Caroline Street North, Waterloo, Ontario N2L 2Y5, Canada}

\emailAdd{lhparker@berkeley.edu}
\abstract{We present the first application of neural field-level baryon-acoustic oscillation (BAO) reconstruction to real spectroscopic survey data. We develop Linear Field Transformer (LiFT), a 3D vision transformer that takes as input the observed galaxy field, its standard reconstruction, and a set of context channels encoding local line of sight, survey coverage, and redshift, and learns to correct standard reconstruction toward the linear density field. We construct a forward-modeling pipeline to produce mock lightcones similar to the DESI Data Release 1 (DR1) luminous red galaxy (LRG) sample, and train LiFT on these. We validate LiFT on held-out simulations, as well as on additional mocks which differ in gravity solver, halo finder, HOD, cosmology, and fiber-assignment history, as well as on mocks analyzed with a distorted distance-redshift relation; ultimately, we find unbiased dilation parameters with consistently tighter constraints than standard reconstruction. Applied to the DESI DR1 LRGs, LiFT improves errors on $\alpha_{\rm iso}$ by $13\%$, $24\%$, and $30\%$ and on $\alpha_{\rm AP}$ by $5\%$, $22\%$, and $30\%$ relative to the DESI DR1 standard reconstruction analysis in the LRG1, LRG2, and LRG3 bins respectively; meanwhile, our DR1 central values remain consistent with DR1 and DR2. This equates to a factor of 1.2, 1.7 and 2.0 increase in Figure of Merit (or effective survey volume) if one were to only use standard reconstruction. Ultimately, these results establish LiFT as a validated, survey-ready tool for current and upcoming galaxy surveys.}
\keywords{large-scale structure, baryon acoustic oscillations, cosmological parameters}
\maketitle
\flushbottom

\section{Introduction}
Baryon acoustic oscillations (BAO) are the imprint on matter clustering of pressure waves from the primordial photon-baryon plasma in the pre-recombination Universe \cite{peebles1970primeval,sunyaev1970,Eisenstein_1998}. Given their characteristic scale, they provide a standard ruler with which to measure the expansion history of the Universe. Since BAO's first detection in galaxy clustering surveys \cite{Eisenstein_2005,Cole_2005}, the BAO method has become a cornerstone of many spectroscopic surveys: from SDSS-III/BOSS \cite{Alam_2017} and SDSS-IV/eBOSS \cite{Alam_2021} to the current Dark Energy Spectroscopic Instrument \cite[DESI;][]{levi2013desiexperimentwhitepapersnowmass,desicollaboration2016desiexperimentisciencetargeting}. At present, the DESI Data Release 1 (DR1) BAO measurements \cite{adame2025desi}, their cosmological interpretation \cite{Adame_2025}, and the subsequent Data Release 2 (DR2) analysis \cite{Abdul_Karim_2025} provide the most precise BAO measurements to date. In combination with the cosmic microwave background (CMB) and type Ia supernovae (SNe), these measurements suggest an expansion history of the Universe that is in tension with $\Lambda$CDM, showing a preference for a time-evolving dark-energy equation of state rather than a fixed cosmological constant at the $\sim2.8$–$4.2\sigma$ level depending on the SNe sample used. Any improvement in BAO distance-measurement precision therefore directly sharpens the constraints on dark energy and on the significance of this deviation from $\Lambda$CDM.

A key factor limiting this precision is non-linear structure formation: large-scale bulk flows displace galaxies from their initial positions, which smears the acoustic feature imprinted upon the galaxy density field \cite{meiksin1999baryonic,seo2008nonlinear}. Standard BAO reconstruction \cite{eisenstein2007improving} reverses this smearing by estimating the displacement field from the observed galaxy density field to linear order using the Zel'dovich approximation \cite{zel1970gravitational}; it then moves galaxies back to their initial positions accordingly, which sharpens the acoustic peak and can improve precision on BAO by tens of percent. However, standard reconstruction is a first-order method that assumes a linear, scale-independent galaxy bias and solves the linearized continuity equation for a curl-free displacement; as a result, it cannot account for non-linear bias or higher-order displacements, which contribute to the residual smearing of the BAO feature. While iterative \cite{schmittfull2017iterative,hada2018iterative}, non-linear \cite{Zhu_2017, Birkin_2018, Shi_2018}, and action-based \cite{Sarpa_2019} methods recover part of this information, a substantial gap between the post-reconstruction field and the true linear field remains. 

To close this gap, \textit{field-level reconstruction} has emerged, whereby one attempts to solve the inverse problem of reconstructing the primordial linear density field from the late-time observed galaxy density field. These methods are motivated by the exceptional robustness of the BAO feature to complex, small-scale physics \cite{angulo2014galaxy, springel2018first, HernandezAguayo2023MillenniumTNG}, and under optimality of the approach would recover all available BAO information. Within field-level reconstruction, two separate research programs have emerged. Explicit field inference aims to explicitly model the likelihood and the prior, and uses high-dimensional optimization \cite{Seljak_2017,Bayer:2022vid,doeser2025learning} or sampling \citep{Jasche_2013,Jasche:2018oym,Schmidt:2018bkr,Schmidt:2020viy,Nguyen:2020hxe,Kostic:2022vok,Bayer:2023rmj,Nguyen:2024yth,Euclid:2024ris,doeser2024bayesian,Simon:2025gwa} to obtain the solutions, relying on differentiable simulators to facilitate this process. However, it suffers from model misspecification issues at the level of the assumed likelihood and forward model, in part due to simplifications needed for differentiable forward models, and its overall computational cost, which can be thousands of forward evaluations per posterior, is often prohibitive at survey scale (see \cite{bayer2026field} for a further discussion). In this work we instead use the so-called implicit field-level inference where the posterior, or its mean, is implicitly learned using a neural network trained on simulated maps of data \citep{shallue2023reconstructing, chen2023effective, floss2024improving, Bottema:2025vww, legin2024posterior, cuesta2024joint, parker2025initial}. Because the simulator in implicit inference is only ever used offline to generate training data, it does not need to be differentiable or admit a tractable likelihood, which makes it easier to use models that more effectively capture small-scale clustering and the halo-galaxy connection  \cite{bayer2026field}. 

Implicit field-level reconstruction has largely been applied in idealized settings, with small-scale periodic simulation boxes applied to dark matter \citep{shallue2023reconstructing, chen2023effective} or, more recently, halos and DESI-like galaxy mocks \cite{parker2025initial, bayer2026field}. However, real survey data introduce various complications that are absent from these tests, such as a non-trivial survey footprint, radial selection, fiber-assignment incompleteness, varying line of sight, and lightcone evolution. To date, no field-level implicit reconstruction method has been applied directly to real observed galaxy data.

\begin{figure}[t]
  \centering
  \includegraphics[width=\textwidth]{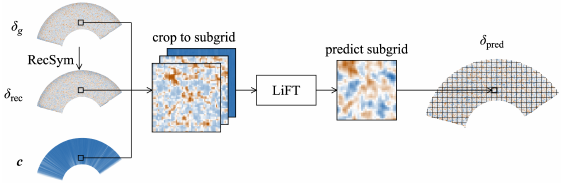}
  \caption{Overview of the LiFT reconstruction pipeline. We feed LiFT subgrids of the observed galaxy overdensity $\delta_g$, its RecSym standard reconstruction field $\delta_{\rm rec}$, and additional context channels that encode position and survey coverage $\bm{c}$. LiFT then estimates a subgrid of the corresponding $z=0$ linear density field. Because standard reconstruction correctly resolves the large-scale features, LiFT only needs to learn local, non-linear corrections. Repeating this process across the survey produces output regions that are then tiled across the volume. Once tiled, we use the power spectrum of the full field for downstream BAO fitting. Two-dimensional slices of just one cap are shown above, however we note that the network operates on three-dimensional subvolumes across both galactic caps in DESI.
  }
  \label{fig:lift-inference-scheme}
\end{figure}

In this work, we introduce Linear Field Transformer (LiFT), an implicit field-level reconstruction method that extends our previous work \cite{parker2025initial, bayer2026field} to realistic survey conditions, and present the first application of field-level BAO reconstruction to real survey data on the DESI DR1 luminous red galaxy (LRG) sample. We utilize LiFT strictly as a reconstruction technique to produce a restored density field on which the BAO is subsequently measured from a Gaussian likelihood analysis of the power spectrum. This allows for a direct comparison with the standard reconstruction pipeline used by surveys like DESI \cite{adame2025desi}. We follow the established DESI analysis practice wherever possible, so that we can isolate the improvements that come from the better reconstruction of initial conditions as opposed to other modeling choices. 

The remainder of this paper is organized as follows. In \cref{sec:obs-and-sims}, we describe the DESI DR1 LRG sample and the forward-modeled simulations used to train and evaluate LiFT. In \cref{sec:recon}, we review standard BAO reconstruction and introduce the LiFT architecture, training procedure, and inference scheme. In \cref{sec:bao}, we present our power-spectrum measurement and BAO-fitting methodology. In \cref{sec:eval}, we validate the method on HalfDome simulations that were not seen during training, as well as on AbacusSummit mocks that use different gravity solver, halo finder, HOD, cosmology, and fiber assignment history from the HalfDome mocks used to train LiFT. We also validate that LiFT is robust to distortions in the redshift-distance relation. We then apply LiFT to the DESI DR1 LRG sample and present the resulting BAO constraints in \cref{sec:desi_results}. Finally, we summarize our conclusions and discuss directions for future work in \cref{sec:conclusions}.

% We utilize field-level inference strictly as a reconstruction technique, producing a restored density field on which the BAO is subsequently measured from a
% Gaussian likelihood analysis of the power spectrum. This allows for a direct comparison with the standard reconstruction pipeline used by surveys like DESI \cite{Adame_2025}. We follow 
% the established DESI analysis practice wherever
% possible, so that we can isolate the improvements that come from the 
% better reconstruction of initial 
% conditions as opposed to other 
% modeling choices. 

\section{Observations and simulations}
\label{sec:obs-and-sims}

\subsection{The DESI DR1 LRG sample}
\label{sec:data:desi}
We analyze the luminous red galaxy (LRG) sample from DESI Data Release~1 (DR1) \citep{adame2025desi}. We analyze all three DESI LRG redshift bins: LRG1 ($0.4 < z < 0.6$), LRG2 ($0.6 < z < 0.8$), and LRG3 ($0.8 < z < 1.1$), which contain $506{,}905$, $771{,}875$, and $859{,}824$ galaxies, respectively \citep{adame2025desi}. The sample is drawn from the DESI DR1 large-scale-structure (LSS) clustering catalogs built on the \texttt{iron} spectroscopic reduction. Specifically, we use \texttt{LSScats/v1.2}, the catalog version used for the DESI DR1 BAO analyses \citep{ross2025construction,adame2025desi}. This catalog includes survey vetoes and matched randoms, together with per-object weights that correct for imaging systematics, redshift failures, and fiber-assignment incompleteness.

\subsection{Forward-modeled mocks}
\label{sec:data:sims}

\begin{figure}[t]
    \centering
    \includegraphics[width=\textwidth]{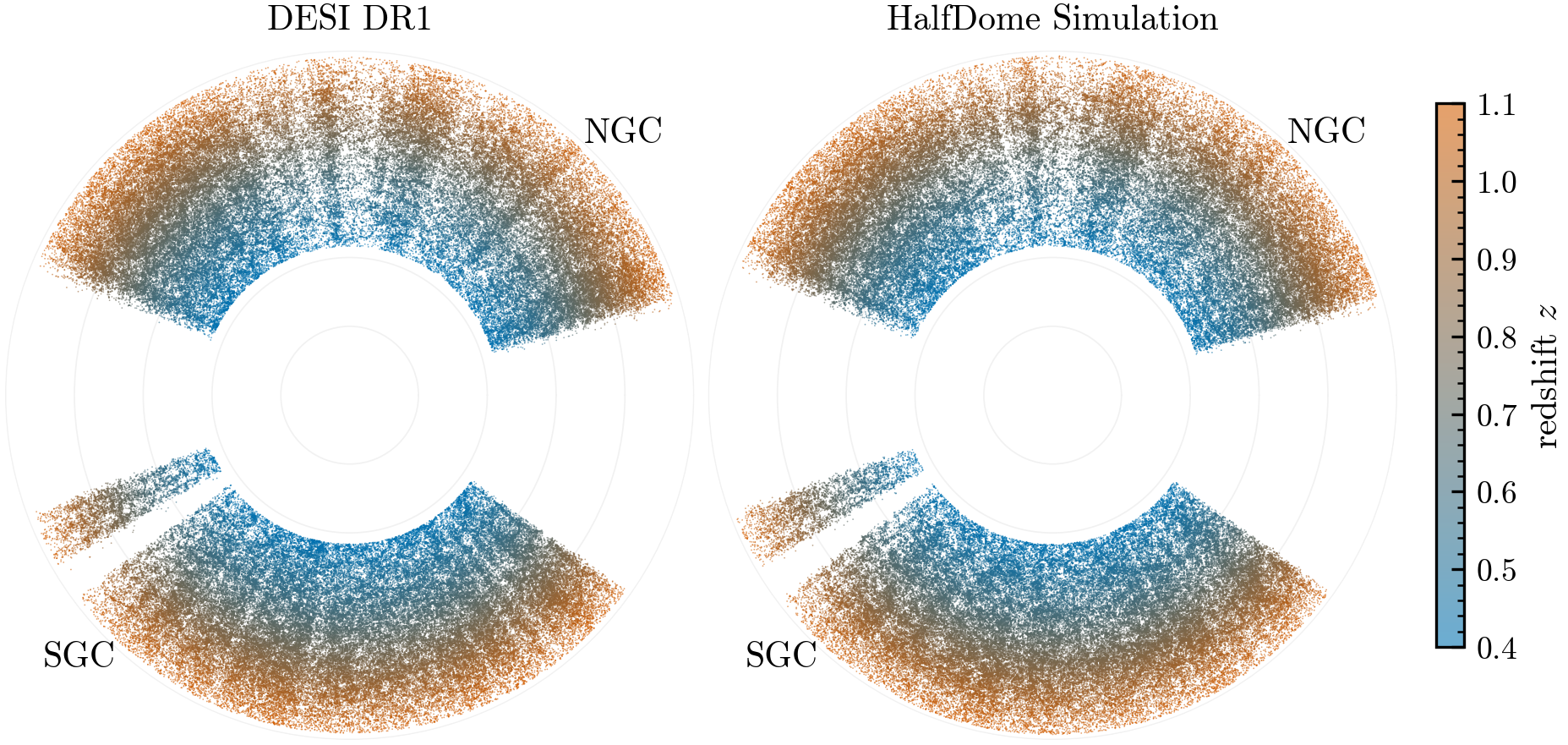}
    \caption{Angular wedge comparison between the DESI DR1 LRG sample (left) and one of our mocks (right). The mocks reproduce the survey footprint, radial selection, and cap geometry of the real data.}
    \label{fig:wedge}
\end{figure}

\subsubsection{HalfDome FastPM simulations}
\label{sec:data:halfdome}
We use the HalfDome suite \citep{bayer2025halfdome}, which consists of full-sky lightcones spanning $0 < z < 4$, simulated at the Planck 2015 cosmology \citep{ade2016planck}. HalfDome provides 11 realizations with different initial linear fields. Each simulation evolves $6144^3$ cold dark matter particles in a cubic box of side $3750\,h^{-1}{\rm Mpc}$ using the FastPM particle-mesh code\footnote{https://github.com/fastpm/fastpm} \cite{feng2016fastpm}. Halos are identified using the Relaxed Friends-of-Friends (RFOF) algorithm \citep{dai2020high}, which is a variant of the standard Friends-of-Friends method. It differs from FoF in that it varies linking length with redshift and halo mass in order to better find low-mass halos. HalfDome has a minimum halo mass of $6 \times 10^{12} M_\odot/h$, and provides halo masses calibrated to the Tinker $M_{200\mathrm{m}}$ mass function \citep{Tinker2008}, consistent with the mass definition adopted for the Tinker bias relation \citep{tinker2010large} used below.
For our samples, we select all halos within the LRG redshift range, namely $0.4 < z < 1.1$.

\subsubsection{Halo Occupation Distribution}
\label{sec:data:mocks}

We populate the HalfDome halos with galaxies to emulate the DESI DR1 LRG sample. Using the standard HOD framework \cite{zheng2007galaxy}, we define the average central and satellite occupations as
\begin{equation}
\begin{aligned}
\langle N_{\rm cen}(M,z)\rangle &= \frac{1}{2}\left[1 + \mathrm{erf}\!\left(
\frac{\log M(z) - \log M_{\rm cut}(z)}{\sqrt{2}\,\sigma_{\log M}(z)}\right)\right], \\
\langle N_{\rm sat}(M,z)\rangle &= \langle N_{\rm cen}(M,z)\rangle
\left(\frac{M(z) - M_0(z)}{M_1(z)}\right)^{\alpha}.
\end{aligned}
\label{eq:hod}
\end{equation}
Motivated by the redshift evolution of the DESI LRG HOD \citep{yuan2024desi}, we specify $\log M_{\rm cut}(z)$ and $\log M_1(z)$ at three anchor redshifts, $z=0.40$, $0.71$, and $1.10$, and interpolate between them using a Piecewise Cubic Hermite Interpolating Polynomial (PCHIP) spline. For each halo, we sample central and satellite occupations from Bernoulli and Poisson distributions, respectively, with the mean occupations defined above. Note that the satellite mean vanishes for $M\leq M_0$. We place centrals at the halo centers, and they inherit their host halo's bulk velocity, while we draw satellites from an NFW phase-space distribution \citep{navarro1997universal}.

For each of the 11 HalfDome realizations, we sample 10 sets of HOD parameters to construct 10 galaxy mocks (see \cref{tab:hod-ranges} for the prior ranges). This exposes the network to plausible uncertainty in the galaxy-halo connection while also acting as data augmentation for the training set. To construct the ensemble, we draw smoothly evolving HOD parameters from Gaussian proposal distributions with separately chosen reference values, informed by the DESI LRG HOD results of \cite{yuan2024desi}. Since those results were obtained using AbacusSummit, we do not impose agreement with their individual best-fit HOD parameters but instead select proposals using the following constraints on the resulting galaxy population:
\begin{enumerate}
    \item Its predicted $n(z)$ exceeds the completeness-corrected DESI LRG $n(z)$ by at least $3\%$ at every redshift\footnote{The condition is one-sided because the catalog is subsequently Bernoulli thinned to the DESI $n(z)$ (\cref{sec:data:selection}), which can only remove galaxies and therefore requires $\bar n_{\rm HOD}(z) \geq \bar n_{\rm DESI}(z)$ everywhere; the $3\%$ margin guards against Poisson fluctuations in the mock $n(z)$ within a $\Delta z = 0.01$ bin.}.  
    \item Its predicted satellite fraction in each of the three LRG bins lies within bands consistent with \cite{yuan2024desi}.
    \item Its predicted linear bias \cite{tinker2010large} matches the value fitted from the DESI DR1 power spectra over $0.03 < k < 0.12\,h\,{\rm Mpc}^{-1}$ to within $10\%$.
\end{enumerate}
Across the eleven realizations, $2,652$ of $14,186$ HOD proposals ($18.7\%$) satisfy the acceptance criteria, among which we select the ones that are furthest from their neighbors so as to span the parameter space. Using this proposal mechanism, the accepted HODs span bin-averaged satellite fractions of $9$-$13\%$, $12$-$16\%$, and $12$-$19\%$, and effective linear biases of $1.84$-$2.04$, $2.05$-$2.22$, and $2.12$-$2.41$, in LRG1, LRG2, and LRG3, respectively. We implement our HOD model using the \textsc{halotools} package \cite{hearin2017forward} within \textsc{nbodykit} \cite{hand2018nbodykit}. We provide the actual ranges of HOD parameters that are sampled and accepted using this scheme in Appendix~\ref{app:hod-ensemble}.

\subsubsection{Selection and thinning}
\label{sec:data:selection}

The DESI DR1 LRG sample's survey geometry includes both radial and angular selection. We account for these effects by thinning our mocks. First, we Bernoulli thin the parent catalog generated by our HOD in $\Delta z =0.01$ redshift intervals by independently retaining each galaxy with $p_{\rm radial}(z)=\bar n_{\rm DESI}(z)/\bar n_{\rm HOD}(z)$ and discarding it otherwise. Because the accepted HOD catalogs are always denser than the target sample, $p_{\rm radial}(z)<1$ for all $z$. This matches our number density to the completeness-corrected DESI number density (and not the true DESI number density).

Next, we account for the variation in coverage over the angular map, produced by the partial DR1 tiling and finite fiber supply. We forward model this effect by depositing the official DESI DR1 randoms onto HEALPix pixels ($N_{\rm side}=512$) and assigning each pixel its fractional sky coverage, estimated from the local random density, multiplied by the mean observed fraction of its randoms. The resulting angular selection probability is 
\begin{equation} 
p_{\rm ang}(\hat{\bm n}) = \min\!\left[1,\,(\bar\Sigma_{\rm ran}\,\Omega_{\rm pix})^{-1}\sum_{r\in P(\hat{\bm n})} f_{{\rm tile},r}\right], 
\label{eq:angular-selection}
\end{equation}
where $P(\hat{\bm n})$ is the HEALPix pixel containing the direction $\hat{\bm n}$, $f_{{\rm tile},r}$ is the local fraction of targets observed at the sky position of random $r$, $\bar\Sigma_{\rm ran}$ is the nominal random surface density summed over the input random catalogs, and $\Omega_{\rm pix}$ is the pixel solid angle. Averaged over the footprint, this factor is $0.923$ and $0.832$ in the NGC and SGC, respectively.

Finally, we account for DESI's mean incompleteness induced by fiber assignment. Specifically, because only some fraction of targets receive an observation on average, we apply a global Bernoulli thinning according to $p_{\rm fiber}=1/\langle w_{\rm comp}\rangle$, where $w_{\rm comp}$ is the DESI completeness weight and the average is taken separately in each cap and redshift bin. This ensures that the mock density matches the observed, non-completeness-corrected DESI sample, and therefore that our mock data has similar shot noise, up to the additional variance induced by the spatial variation of the DESI incompleteness weights. We note that the real fiber assignment process contains a density dependence on its incompleteness, whereas here we model it only on average; we show in \cref{sec:abacus-altmtl} that this simplification nonetheless does not materially affect our BAO constraints. We evaluate $p_{\rm fiber}$ separately in each cap and each of the three analysis redshift bins, $0.4\leq z<0.6$, $0.6\leq z<0.8$, and $0.8\leq z<1.1$, obtaining $(0.7935,\,0.7809,\,0.7892)$ in the NGC and $(0.7058,\,0.6911,\,0.7027)$ in the SGC, respectively. Combined with the $p_{\rm ang}$ factor above, this results in a factor of $0.692$ thinning over the full sample, reproducing the $69.2\%$ DR1 LRG assignment completeness \cite{adame2025desi}.

\begin{figure}[t]
    \centering
    \includegraphics[width=0.65\textwidth]{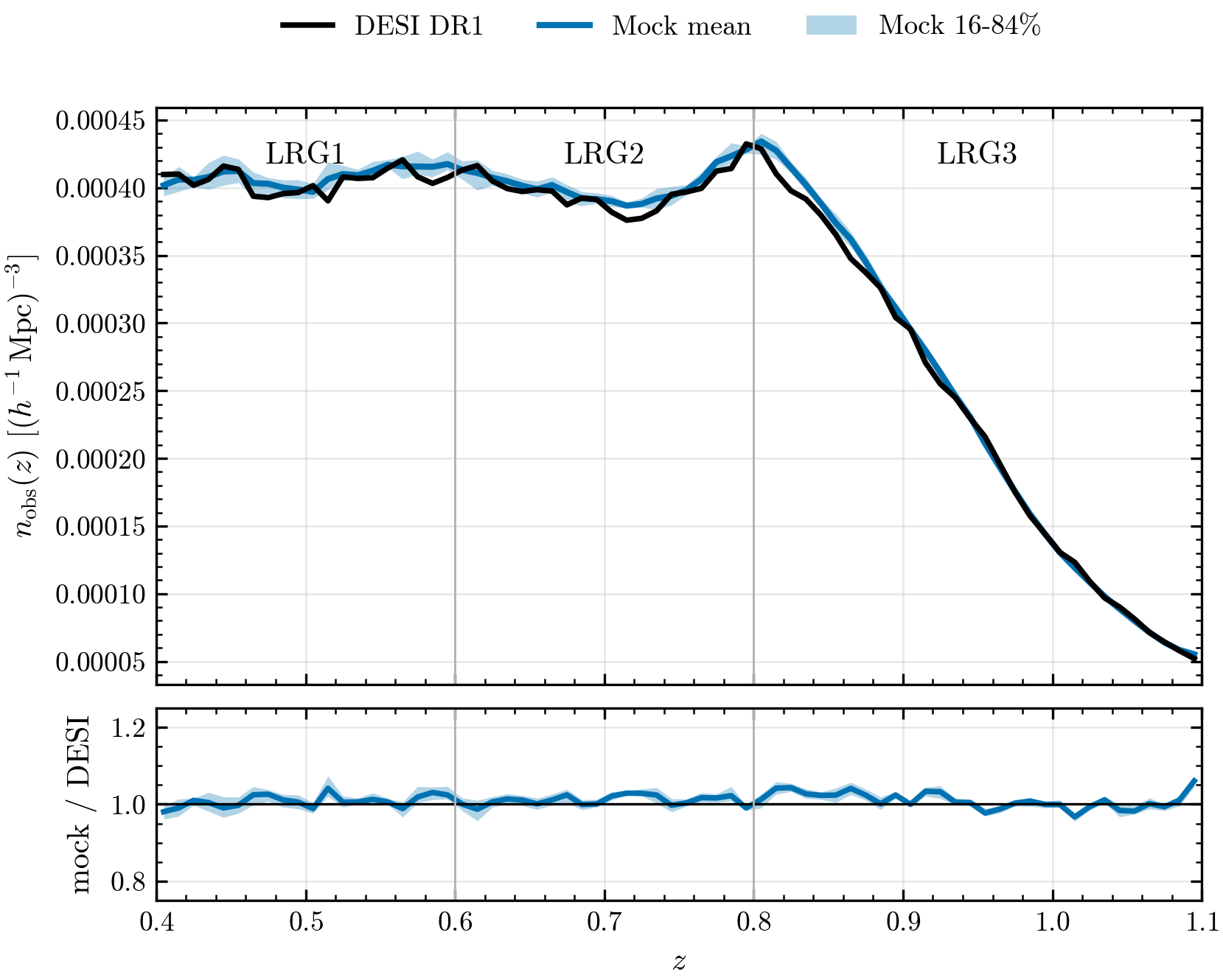}
    \caption{Comoving number density over 10 HOD draws of one of our forward-modeled HalfDome mocks (blue), compared with the DESI DR1 LRG sample (black), over the full LRG range. The DESI curve is the official completeness-corrected $n(z)$, and the mock galaxies are weighted by their compensating fiber-assignment weights. The lower panel is the ratio of the mock $n(z)$ to the DESI $n(z)$, showing agreement at the few-percent level.}
    \label{fig:nz}
\end{figure}

Ultimately, this procedure reproduces the mean radial and angular selection of the survey at the few-percent level across the LRG redshift range, as shown in \cref{fig:nz}, where we compare the weighted number density of our mocks (weighted by $1/p_{\rm fiber}$) with the completeness-corrected DESI $n(z)$. 

\subsubsection{Random catalogs}
\label{sec:data:randoms}

The angular selection of the survey is identical between DESI and our mocks by construction. However, because of stochasticity in the thinning stages, the radial selection slightly differs. Therefore, while we retain the angular positions of the official DESI randoms, we reassign each random the redshift of a mock galaxy drawn at random from the same imaging region and a weight equal to the mean observed fraction of  \cref{sec:data:selection} evaluated at its own sky pixel, after which we normalize the total random weight to the total galaxy weight separately within each imaging region. Ultimately, this process mirrors the DESI randoms treatment, and ensures that there is no mismatch between the redshift distributions of a mock and its corresponding randoms.

\subsubsection{Galaxy density fields}
\label{sec:data:density-fields}

\begin{figure}[t]
    \centering
    \includegraphics[width=0.8\textwidth]{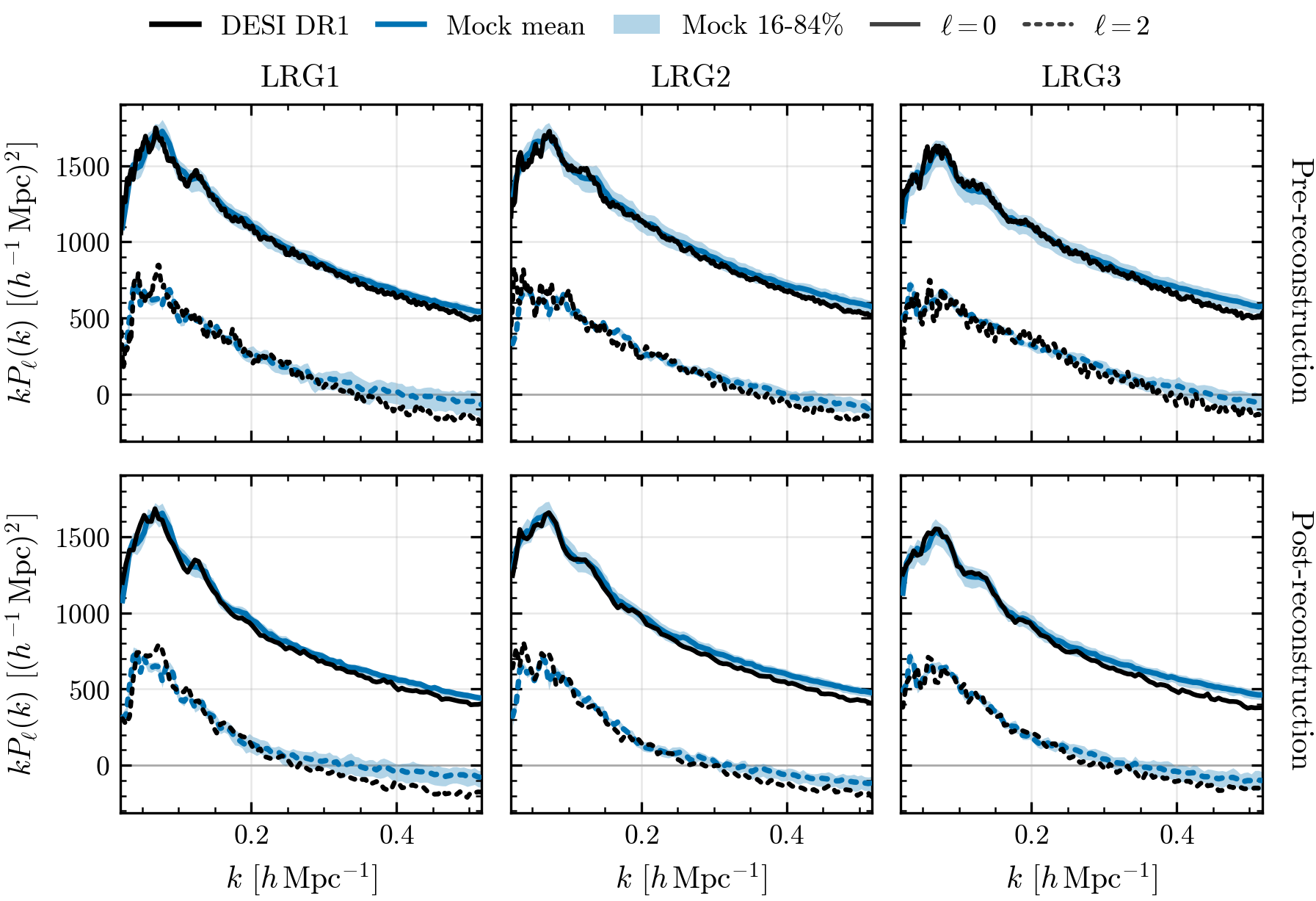}
    \caption{Power-spectrum multipoles of the DESI DR1 LRG sample (black) compared with our forward-modeled mocks (blue), before (top) and after (bottom) standard BAO reconstruction. The solid and dashed blue lines show the monopole and quadrupole means and the shaded bands the 16-84\% spreads over one HalfDome seed's 10 HOD realizations. For details on BAO reconstruction, see \cref{sec:standard-reconstruction}.}
    \label{fig:pk-validation}
\end{figure}

For LiFT, we use the galaxy overdensity field on a mesh for both our mocks and the real DESI data. To create the overdensity field, we combine the NGC and SGC for all LRG bins and deposit the weighted galaxies, $D(\bm{x})$, and their corresponding randoms, $R(\bm{x})$, on a mesh using triangular-shaped-cloud (TSC) assignment with interlacing. We define the galaxy overdensity then as 
\begin{equation}
    \delta_{\rm g}(\bm{x})
    = \frac{D(\bm{x})}{\alpha_{\rm norm}R(\bm{x})}-1,
    \qquad
    \alpha_{\rm norm}
    = \frac{\sum_i w_{{\rm g},i}}{\sum_j w_{{\rm r},j}}.
    \label{eq:galaxy-overdensity}
\end{equation}
We use an observer-centered Cartesian grid with $1024^3$ cells and a cell size of $5.21\,h^{-1}{\rm Mpc}$, corresponding to a Nyquist frequency of $k_{\rm nyq} = 0.6\,h{\rm Mpc}^{-1}$. To avoid unstable values in poorly sampled cells near the survey boundary, we retain only cells within the footprint whose random density exceeds $5\%$ of its mean within the footprint. 

\subsubsection{Linear fields}
\label{sec:data:linear-fields}
Our method learns to predict the linear density field of each HalfDome realization. Because in linear theory the density contrast is just the initial Gaussian field rescaled by the growth factor, we set 
\begin{equation}
    \delta_{\rm lin}(\bm{x}) \equiv \delta_{\rm lin}(\bm{x}, z=0)
    = \frac{D(0)}{D(z_{\rm init})}\,\delta_{\rm lin}(\bm{x}, z_{\rm init}),
    \label{eq:linear-target}
\end{equation}
which is the initial conditions of the simulation evolved to $z=0$ in real space and without redshift-space distortions. Normalizing to $z=0$ is just a rescaling convention, since the BAO scale is invariant under the scalar rescaling by $D(z)$. 

In practice, FastPM provides the linear field of each realization in Fourier space on the $6144^3$ grid of the $3750\,h^{-1}{\rm Mpc}$ HalfDome box. We truncate this to the Nyquist frequency of our galaxy density field, $k_{\rm nyq}=0.6\,h\,{\rm Mpc}^{-1}$, and inverse Fourier transform, which yields the same cell size as the observer-centered grid in \cref{sec:data:density-fields}. Because the HalfDome lightcones are generated by periodically replicating the simulation box, we consistently tile this periodic mesh onto the observer-centered grid of \cref{sec:data:density-fields}, creating a linear target for our network that is aligned cell-for-cell with the galaxy fields.

\section{Reconstruction methods}
\label{sec:recon}

\subsection{Standard BAO reconstruction}
\label{sec:standard-reconstruction}

Standard BAO reconstruction \cite{eisenstein2007improving} is commonly used to partially reverse the broadening of the acoustic feature caused by bulk flows and traced by the observed galaxy field. It does so by using the Zel'dovich approximation \citep{zel1970gravitational} to estimate the displacement field, after which it moves observed tracers backwards. This has been shown to materially sharpen the BAO feature and partially restore the correlation with the initial density field \citep{eisenstein2007improving}. As LiFT is a correction to standard reconstruction, we provide an overview below.

Let $\bm{x}$ be the Eulerian position of a fluid element; it is related to its Lagrangian coordinate $\bm{q}$ by $\bm{x}=\bm{q}+\bm{\Psi}$, where $\bm{\Psi}$ is the Lagrangian displacement field. Mass conservation gives $\delta_{\rm lin}= -\nabla\cdot\bm{\Psi}$, and the growing-mode, curl-free solution at linear order is therefore given by
\begin{equation}
    \bm{\Psi}(\bm{x})
    =
    -\nabla\nabla^{-2}\delta_{\rm lin}(\bm{x}).
    \label{eq:zeldovich-displacement}
\end{equation}

For a linearly biased tracer in real space, $\delta_g=b\,\delta_{\rm lin}$, so that $\nabla\cdot\bm{\Psi}=-\delta_{\rm g}/b$. In redshift space, the line-of-sight peculiar velocity produces an additional displacement $f(\bm{\Psi}\cdot\widehat{\bm r})\widehat{\bm r}$. The corresponding linearized continuity equation is therefore
\begin{equation}
\bm{\nabla}\cdot\bm{\Psi}
+ \beta\,\bm{\nabla}\cdot
\left[
    \left(\bm{\Psi}\cdot\widehat{\bm r}\right)
    \widehat{\bm r}
\right]
=
-\frac{\delta_{\rm g}}{b}.    \label{eq:recon-rsd}
\end{equation}
where $\beta\equiv f/b$, $f=d\ln D/d\ln a$ is the logarithmic linear growth rate, and $\widehat{\bm r}$ is the local line of sight. 

We solve \cref{eq:recon-rsd} using the iterative Fourier-space method \textsc{IterativeFFT} \citep{burden2015reconstruction}, implemented in \textsc{pyrecon}\footnote{https://github.com/cosmodesi/pyrecon}. At first order, the displacement is curl-free and can be written as $\bm{\Psi}=\nabla\phi$. Although the line-of-sight field $(\bm{\Psi}\cdot\widehat{\bm r})\widehat{\bm r}$ has non-zero curl, its solenoidal component has zero divergence and does not contribute to \cref{eq:recon-rsd}. The potential may consequently be updated according to
\begin{equation}
    \nabla^2\phi^{(n+1)}(\bm{x})
    =
    -\frac{\delta_{\rm g}(\bm{x})}{b}
    -\beta\,\bm{\nabla}\!\cdot
    \left[
        \left(\bm{\nabla}\phi^{(n)}(\bm{x})\!\cdot\!
        \widehat{\bm r}(\bm{x})\right)
        \widehat{\bm r}(\bm{x})
    \right],
    \qquad
    \nabla^2\phi^{(0)} = -\delta_{\rm g}/b.
    \label{eq:iterative-fft}
\end{equation}
Here, we evaluate gradients and divergences using FFTs at each step. After the iterations are performed, the resulting source is inverse-Laplacian filtered to obtain $\phi^{(n+1)}$, and consequently $\bm{\Psi}$.

Once $\bm{\Psi}$ has been computed, we adopt the symmetric reconstruction convention (RecSym), in which both galaxies and randoms are shifted by $-\bm{\Psi}$. Because the same redshift-space shift is applied to both catalogs, RecSym preserves large-scale redshift-space distortions and is less sensitive to errors in the estimated displacement than the alternative RecIso convention \citep{White2015:1504.03677,Chen:2024tfp,chen2026extensive,Paillas_2025}. 

To produce the reconstructed fields for both the DESI data and our mocks, we follow the DESI conventions \cite{adame2025desi}. Specifically, we reconstruct the full $0.4< z<1.1$ LRG sample at once, but perform reconstruction in the NGC and SGC independently. For both real data and mocks, we use $b=2.0$ and $f=0.83$. We use three iterations of \textsc{IterativeFFT}, and during the FFTs, we deposit particles and randoms on a grid with $4\,h^{-1}{\rm Mpc}$ cells using cloud-in-cell assignment. Additionally, to remove non-linear, small-scale structure, we smooth the density field $\delta_{\rm g}(\bm k) \to \mathcal{S}(k)\delta_{\rm g}(\bm k)$ with $\mathcal{S}(k)=\exp(-k^2R_{\rm sm}^2/2)$, and $R_{\rm sm}=15h^{-1} \rm Mpc$, to suppress nonlinear small-scale modes.

\subsection{LiFT: Linear Field Transformer}
In this work, we introduce LiFT (Linear Field Transformer), a neural-network-based method that recovers the linear density field directly from the observed galaxy survey as a correction to standard reconstruction. While standard reconstruction sharpens the acoustic feature relative to the raw galaxy density field, it estimates displacements using a linearly biased, first-order approximation and as a result does not account for the nonlinear galaxy-matter relation or the higher-order displacements that smear the acoustic feature. To overcome these limitations, we train a neural network to recover the true linear field from the mocks described above, in which nonlinear bias and nonlinear dynamics are present and the true linear field is known:
\begin{equation}
    \textrm{LiFT}:\;
    \bigl(
        \delta_g,\;
        \delta_{\rm rec},\;
        \bm{c}
    \bigr)
    \;\longmapsto\;
    \delta^{\rm pred}_{\rm lin},
    \label{eq:transformer-map}
\end{equation}
where $\delta_g$ is the observed galaxy overdensity, $\delta_{\rm rec}$ is the reconstructed field, $\bm{c}$ denotes a set of context channels encoding the local survey geometry, and $\delta_{\rm lin}$ is the target $z=0$ linear field.

\subsubsection{Model architecture}
\label{sec:architecture}
To learn this mapping, we previously trained a convolutional neural network (CNN) on periodic cubic boxes with a single, global line of sight  \cite{parker2025initial,bayer2026field}. CNNs rely on convolutions of kernels, an operation which is translation equivariant, applying identical weights with a fixed orientation everywhere in the volume. This symmetry works for global lines of sight, but is broken when the redshift-space distortions follow a varying line of sight. To address this, we instead use a three-dimensional vision transformer \cite{vaswani2017attention,dosovitskiy2020image} supplied with the local line of sight as an input channel, which is able to learn a correction that accounts for the rotating line of sight. To that end, we pass the transformer the three components of the normalized cell position $\bm{x}/r_{\rm norm}$, where $\bm{x}$ is the observer-centered comoving position of the cell and $r_{\rm norm} = \chi(z=1.1)$ is the comoving distance to the far edge of the LRG sample, which enables the network to account for the local line of sight as well as the redshift of each cell. In addition, we also provide the angular selection map $p_{\rm ang}(\hat{\bm n})$ from \cref{sec:data:selection} projected onto the analysis grid as the fourth channel in $\bm{c}$; this enables the network to account for local survey geometry and masked regions.

During network training and inference, we operate on subsets of the full survey volume, as in \cite{parker2025initial}. Training on subvolumes is possible because the large-scale modes are already accurately modeled by standard reconstruction, meaning the network only needs to learn local, small-scale corrections. Specifically, we feed the network cubic input regions $\Omega$ spanning $50^3$ cells ($260\,h^{-1}{\rm Mpc}$), and predict subvolumes of the linear field $\Omega'$ spanning $32^3$ cells ($167\,h^{-1}{\rm Mpc}$). This ensures that every output cell is supported by a buffer of at least $9$ cells ($47\,h^{-1}{\rm Mpc}$) of surrounding context. We choose these scales such that both input and output comfortably exceed the BAO scale (${\sim}100\,h^{-1}{\rm Mpc}$). This provides an advantage both in terms of compute (by reducing the activation size of the network) and also significantly increases the training set size by partitioning one seed into many subvolumes; indeed, the DR1 survey volume of a single lightcone contains ${\sim}2\times10^{3}$ non-overlapping predicted regions.

The transformer itself is designed as a 3D vision transformer. It partitions the $6$-channel, $50^3$-cell input into non-overlapping patches of $5^3$ cells, giving $(50/5)^3 = 10^3$ patches; each patch is flattened to its $6\times5^3 = 750$ values and linearly projected to a token of embedding dimension $256$. These $10^3$ tokens are processed by eight pre-norm transformer blocks with eight-head self-attention, augmented by a learned three-dimensional relative-position bias \citep{shaw2018self,liu2021swin}, after which the tokens are reassembled onto the spatial grid and upsampled using a transposed convolution followed by three residual convolutional blocks, and a final projection to the single output channel. The prediction is then center-cropped to $32^3$ cells. Ultimately, this results in roughly $9.3$ million trainable parameters. For a review of transformer architectures in an astrophysical context, we direct the reader to Appendix A3 of \cite{parker2024astroclip}.

\subsubsection{Training}
\label{sec:training}
We train a single network over the full $0.4<z<1.1$ LRG volume, rather than a separate network for each LRG bin: the three LRG bins are only separated at the power-spectrum stage (\cref{sec:pk}). We train the network using a standard mean squared error (MSE) loss, with additional weighting to account for the survey mask and coverage. Specifically, we use
\begin{equation}
    \mathcal{L}
    =
    \frac{\sum_{\bm{x}\in\Omega'} w(\bm{x})
    \left[\delta^{\rm pred}_{\rm lin}(\bm{x})-\delta_{\rm lin}(\bm{x})\right]^2}
    {\sum_{\bm{x}\in\Omega'} w(\bm{x})},
    \label{eq:training-loss}
\end{equation}
where $w(\bm{x}) = M(\bm{x})\,p_{\rm ang}(\hat{\bm n})$ and $M$ is the survey mask that sets regions outside of the mask to zero, and $p_{\rm ang}(\hat{\bm n})$ is the coverage map defined in \cref{sec:data:selection} that downweights poorly observed regions. Because the loss is a (weighted) MSE, the network's optimum is the conditional mean of the linear field given its inputs; we characterize the consequences of this choice for the two-point statistics of the reconstruction in \cref{app:mse}.

During training, for each seed, we draw subvolumes at random from the full grid, and then reject positions for which $<10\%$ of $\Omega'$ lies within $M$. For each epoch, we draw $5000$ subvolumes evenly from a pool of mocks that rotates the $10$ HOD realizations. We perform training with the AdamW \cite{loshchilov2019decoupledweightdecayregularization} optimizer using a warmup-plus-cosine decay schedule, using a peak learning rate of $3\times10^{-4}$, a linear warmup over the first $3$ epochs, a weight decay of $0.01$, and a batch size of $32$ subvolumes, for $60$ epochs in total. Ultimately, because of the low total number of HalfDome seeds available, we train three networks in a $3$-fold cross-validation strategy over the eleven HalfDome seeds ($4$/$4$/$3$ held out); the network applied to the DESI data is trained on all eleven. Note that a single network is trained and applied over the full $0.4<z<1.1$ LRG volume rather than one per redshift bin.

\subsubsection{Inference}
\label{sec:inference}
Once trained, the network produces a prediction of the linear field of size $\Omega' = 167\,h^{-1}{\rm Mpc}$ (see above). In order to infer the full survey volume, we slide the trained network across the analysis grid. Let $\{\bm{x}_i\}$ denote the centers of the predicted regions, placed on a regular lattice of spacing equal to the side of $\Omega'$, such that the regions $\Omega'_i$ tile the $1024^3$ grid exactly. Denoting the collected input fields by $I = (\delta_g, \delta_{\rm rec}, \bm{c})$, the full-volume prediction is then given by
\begin{equation}
    \delta^{\rm pred}_{\rm lin}
    =
    \textrm{LiFT} \circledast I
    \;\equiv\;
    \sum_i
    \mathbb{1}_{\Omega'_i}\,
    \textrm{LiFT}\!\left[\,I_{\Omega_i}\right],
    \label{eq:inference-tiling}
\end{equation}
where $\textrm{LiFT}[\,\cdot\,]$ denotes the network of \cref{eq:transformer-map} evaluated on the input subvolume $\Omega_i$ and $\mathbb{1}_{\Omega'_i}$ is the indicator function of the predicted region. We note that $\circledast$ denotes this strided, patchwise application rather than a true convolution. We perform this inference scheme both for the HalfDome/AbacusSummit mocks as well as for the real DESI data.

\subsection{Reconstruction performance}
\label{sec:reconstruction-performance}
\begin{figure}[t]
    \centering
    \includegraphics[width=0.8\textwidth]{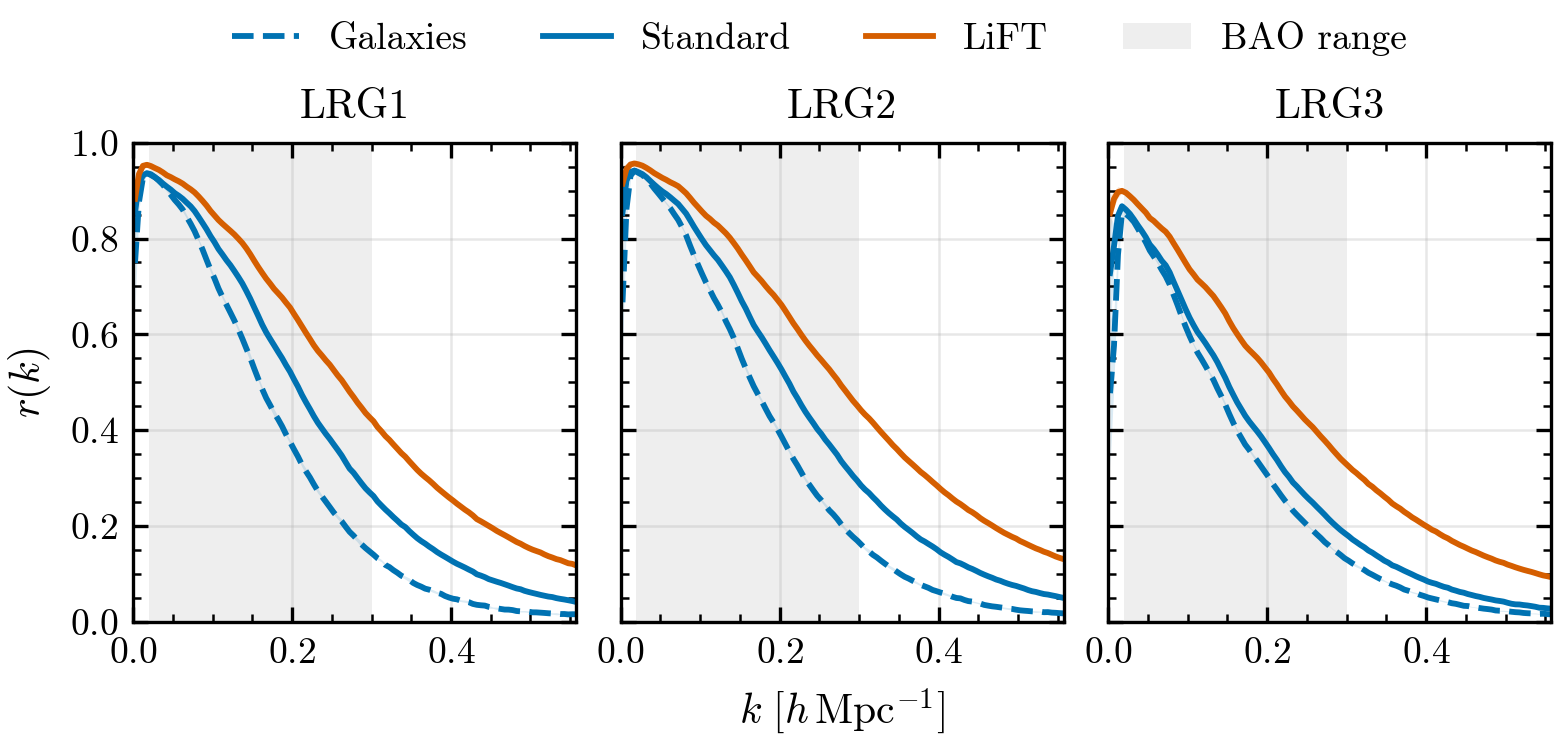}
    \caption{Cross-correlation with the $z=0$ linear field for the raw galaxies, standard reconstruction, and the LiFT prediction. All curves are averaged over the same eleven HalfDome realizations; the LiFT predictions are pooled from the three held-out cross-validation folds. The shaded region marks the $k$-range used for BAO fitting.}
    \label{fig:reconstruction-correlation}
\end{figure}

Before fitting the BAO, we evaluate the network's performance on the HalfDome simulations using $3$-fold cross-validation described in \cref{sec:training}. To that end, we compute the cross-correlation coefficient between each field and the known true linear field within the masked region as 
\begin{equation}
    r(k)
    \equiv
    \frac{P_{{\rm pred},{\rm lin}}(k)}
    {\left[P_{{\rm pred},{\rm pred}}(k)P_{{\rm lin},{\rm lin}}(k)\right]^{1/2}}.
    \label{eq:reconstruction-correlation}
\end{equation}
We average $r(k)$ across the eleven inferred fields, and present the average in \cref{fig:reconstruction-correlation}. Additionally, we compute the same cross-correlation between standard reconstruction ($\delta_{\rm rec}$) and the linear field, as well as the raw galaxy field and the linear field. We observe that LiFT is clearly more strongly correlated with the linear field throughout the BAO range in all three redshift bins. 

\section{BAO fitting}
\label{sec:bao}

\subsection{Two-point statistics}
\label{sec:pk}
To fit the BAO, we compress the reconstructed fields into power-spectrum multipoles. The measured wavevectors are expressed in the coordinate system defined by the fiducial cosmology, and we denote their magnitude and line-of-sight cosine by $(k,\mu)$. Their mapping to the dilated template coordinates $(k',\mu')$ is described in \cref{sec:bao-dilation}. For the LiFT prediction, we cut out the comoving shell corresponding to each LRG bin, set everything outside of the shell and the mask to zero, and Fourier transform the resulting field directly with \textsc{pypower}\footnote{https://github.com/cosmodesi/pypower}, using local (first-point) line of sight \cite{yamamoto2006measurement}. Because the field is defined natively on the grid, we do not use any random catalog, assignment-kernel compensation, or shot-noise subtraction. We measure the power spectrum multipoles $\ell = 0, 2, 4$ separately in each Galactic cap. We then combine the caps, weighted by their normalizations. We compute the power spectra in bins of width $\Delta k = 0.005\,h\,{\rm Mpc}^{-1}$. While only the power spectrum monopole and quadrupole are used for the BAO fits, the hexadecapole is needed to compute the covariance in \cref{sec:cov}. For the standard-reconstruction results below, we measure the same power spectrum multipoles from the shifted galaxies and randoms with FKP weights applied \citep{feldman1993power}. We follow DESI DR1 \cite{adame2025desi}, and deposit both on a $1168^3$ mesh of box size $7008\,h^{-1}{\rm Mpc}$ ($6\,h^{-1}{\rm Mpc}$ cells) with TSC assignment and interlacing.

\subsection{BAO dilation parameters}
\label{sec:bao-dilation}
When converting the observed angular positions and redshifts of galaxies into comoving coordinates using a fiducial cosmology, the true distance-redshift relation may differ from the fiducial one. As a result, the inferred clustering scales along both the radial and transverse directions can be dilated. These dilations are described as
\begin{equation}
    \alpha_\parallel(z)
    =
    \frac{D_H(z)/r_d}{D_H^{\rm fid}(z)/r_d^{\rm fid}},
    \qquad
    \alpha_\perp(z)
    =
    \frac{D_M(z)/r_d}{D_M^{\rm fid}(z)/r_d^{\rm fid}},
    \label{eq:alpha-par-perp}
\end{equation}
where $D_M(z)$ is the transverse comoving distance, $D_H(z)=c/H(z)$, and $r_d$ is the sound horizon at the drag epoch. In this analysis, we rewrite the dilation parameters in an equivalent basis, given by
\begin{equation}
    \alpha_{\rm iso}
    =
    \left(\alpha_\parallel\alpha_\perp^2\right)^{1/3},
    \qquad
    \alpha_{\rm AP}
    =
    \frac{\alpha_\parallel}{\alpha_\perp},
    \label{eq:alpha-iso-ap}
\end{equation}
where $\alpha_{\rm iso}$ can be thought of as the volume-averaged isotropic dilation of the BAO scale and $\alpha_{\rm AP}$ is the anisotropic distortion that exists between the radial and transverse directions to the line of sight. Here, the latter is also known as the Alcock-Paczynski dilation. 

At the level of the power spectrum, we write the Fourier modes measured in the fiducial coordinate system as $(k,\mu)$ (with $\mu$ the cosine of the angle between the wavevector and the line of sight), and then write the corresponding coordinates at which the dilated BAO template is evaluated, which we denote with $(k',\mu')$, by defining the radial and transverse components of the $k$-modes as 
\begin{equation}
    k'_\parallel = \frac{k\mu}{\alpha_\parallel},
    \qquad
    k'_\perp = \frac{k\sqrt{1-\mu^2}}{\alpha_\perp},
\end{equation}
which gives
\begin{align}
    k'
    &=
    \frac{k\,\alpha_{\rm AP}^{1/3}}{\alpha_{\rm iso}}
    \left[1+\mu^2\left(\alpha_{\rm AP}^{-2}-1\right)\right]^{1/2},
    \label{eq:k-prime}\\
    \mu'
    &=
    \frac{\mu}{\alpha_{\rm AP}}
    \left[1+\mu^2\left(\alpha_{\rm AP}^{-2}-1\right)\right]^{-1/2}.
    \label{eq:mu-prime}
\end{align}
Thus, $(k',\mu')=(k,\mu)$ when $\alpha_{\rm iso}=\alpha_{\rm AP}=1$.

\subsection{Modeling the LiFT power spectrum}
\label{sec:LiFT-model}
By design, LiFT is an MSE-trained estimator of a Gaussian field. As a result, its optimum is the conditional mean of the linear field given the inputs, which has Wiener-filter-like two-point statistics with any nonlinear residual absorbed into $n$ (we derive this in \cref{app:mse}). Therefore, we model its prediction as the true linear field filtered by an anisotropic transfer function plus additive reconstruction noise uncorrelated with $\delta_{\rm lin}$,
\begin{equation}
    \delta^{\rm pred}_{\rm lin}(\bm{k})
    =
    G(k,\mu)\,\delta_{\rm lin}(\bm{k})
    +
    n(\bm{k}),
    \label{eq:LiFT-transfer-noise}
\end{equation}
so that the power spectrum is $G^2 P_{\rm lin} + P_n$. A direct consequence of modeling the LiFT reconstruction this way is that a single response $B \equiv G^2$ multiplies both the wiggle and nowiggle components. This differs from the standard BAO model \cite{Chen:2024tfp}, in which the acoustic feature is damped separately from the broadband to account for nonlinear evolution \cite{eisenstein2007robustness,seo2008nonlinear}, which standard reconstruction partially undoes but does not eliminate, therefore requiring the asymmetric treatment to persist. For LiFT, by contrast, $G$ is the response of a filter applied to the field rather than a smearing of galaxy pairs, and it suppresses smooth and oscillatory power alike. As a result, there is no need to introduce an additional damping of wiggle relative to no wiggle components of the initial field, so the damping of both is parametrized by the same term.  Therefore, we write our network's template model as 
\begin{equation}
    P(k,\mu)
    =
    B(k',\mu')\,(P_{\rm nw}(k') +
    \,P_{\rm w}(k'))
    +
    D(k,\mu),
    \label{eq:LiFT-model}
\end{equation}
where $P_{\rm nw}$ and $P_{\rm w}$ are the smooth and wiggle components of the $z=0$ linear template, $(k',\mu')$ are the AP-dilated coordinates parameterized by $(\alpha_{\rm iso}, \alpha_{\rm AP})$, and $D(k,\mu)$ is the cubic-spline basis in DESI's standard model that absorbs additive uncorrelated reconstruction noise. Despite this treatment, we nonetheless also fit the DESI DR1 LiFT reconstruction with a template in which the smooth component is given its own damping in \cref{app:untied}; ultimately, we find that tying the two responses has negligible impact on the dilation parameters, their errors, and the best-fit $\chi^2$.

We parameterize $B(k,\mu)$ (where $(k', \mu')$ is now implied) in the above model as
\begin{equation}
    B(k,\mu)
    =
    \left(b_0 + b_2\,\mu^2\right)^2
    \exp\!\left[
        -\tfrac{1}{2}k^2
        \left(\mu^2\Sigma_\parallel^2 + (1-\mu^2)\Sigma_\perp^2\right)
    \right].
    \label{eq:LiFT-response}
\end{equation}
Here, we have adopted the same form for $B$ as the DESI DR1 template, except that we have exchanged the naming of the galaxy bias and growth rate $(b,f)$ for $b_0, b_2$. We make the choice to keep the same form as DESI because the form is flexible enough to describe the measured network propagator (see \cref{fig:propagator-fits}). We change the names to clarify that, while the functional form is shared, one should not interpret it as accounting for the physical effects in galaxies, but rather as describing imperfections in LiFT's reconstruction of the linear field. Specifically, the Gaussian damping is the roll-off of the network transfer, which suppresses poorly constrained small-scale modes, rather than a model of nonlinear smearing. The anisotropic amplitude $b_2\,\mu^2$ is not a Kaiser factor (LiFT removes redshift-space distortions by construction) but instead a way of accounting for the residual line-of-sight anisotropy of the transfer.

\subsection{Priors}
\label{sec:priors}

For LiFT, we calibrate the priors on $(b_0, b_2, \Sigma_\parallel, \Sigma_\perp)$ using the full-sky HalfDome simulations in each LRG redshift bin. Specifically, we compute the propagator
\begin{equation}
  G(k,\mu)
  \equiv
  \frac{P_{\rm pred,lin}(k,\mu)}{P_{\rm lin}(k)}
  =
  \left(b_0 + b_2\,\mu^2\right)
  \exp\!\left[
      -\tfrac{1}{4}k^2
      \left(\mu^2\Sigma_\parallel^2 + (1-\mu^2)\Sigma_\perp^2\right)
  \right],
  \label{eq:LiFT-propagator}
\end{equation}
such that $B=G^2$, with $P_{\rm pred,lin}$ the cross-spectrum between the LiFT prediction and the true linear field. 

We compute $G(k,\mu)$ using the training seeds of the 3-fold cross-validation networks. We use training seeds rather than held-out inference seeds in order to ensure that at inference time, the held-out simulations are free of methodological choices. Once computed, for each LRG bin, we average $G(k,\mu)$ over the training seeds, and fit \cref{eq:LiFT-propagator} over the BAO $k$-range, parameterizing $(b_0, b_2, \Sigma_\parallel, \Sigma_\perp)$. Ultimately, this strategy is similar in spirit to the simulation-informed priors of our previous work \cite{bayer2026field}; however, here the propagator is measured against the true linear field rather than fixed to the maximum a posteriori values of a BAO fit to the mean of 900 periodic-box mocks, and we also include an additional anisotropic $b_2\mu^2$ term required by the varying line of sight. The calibrated values then serve as the centers of Gaussian priors with the DESI widths rather than as fixed parameters.

\begin{figure}[t]
\centering
\includegraphics[width=0.8\textwidth]{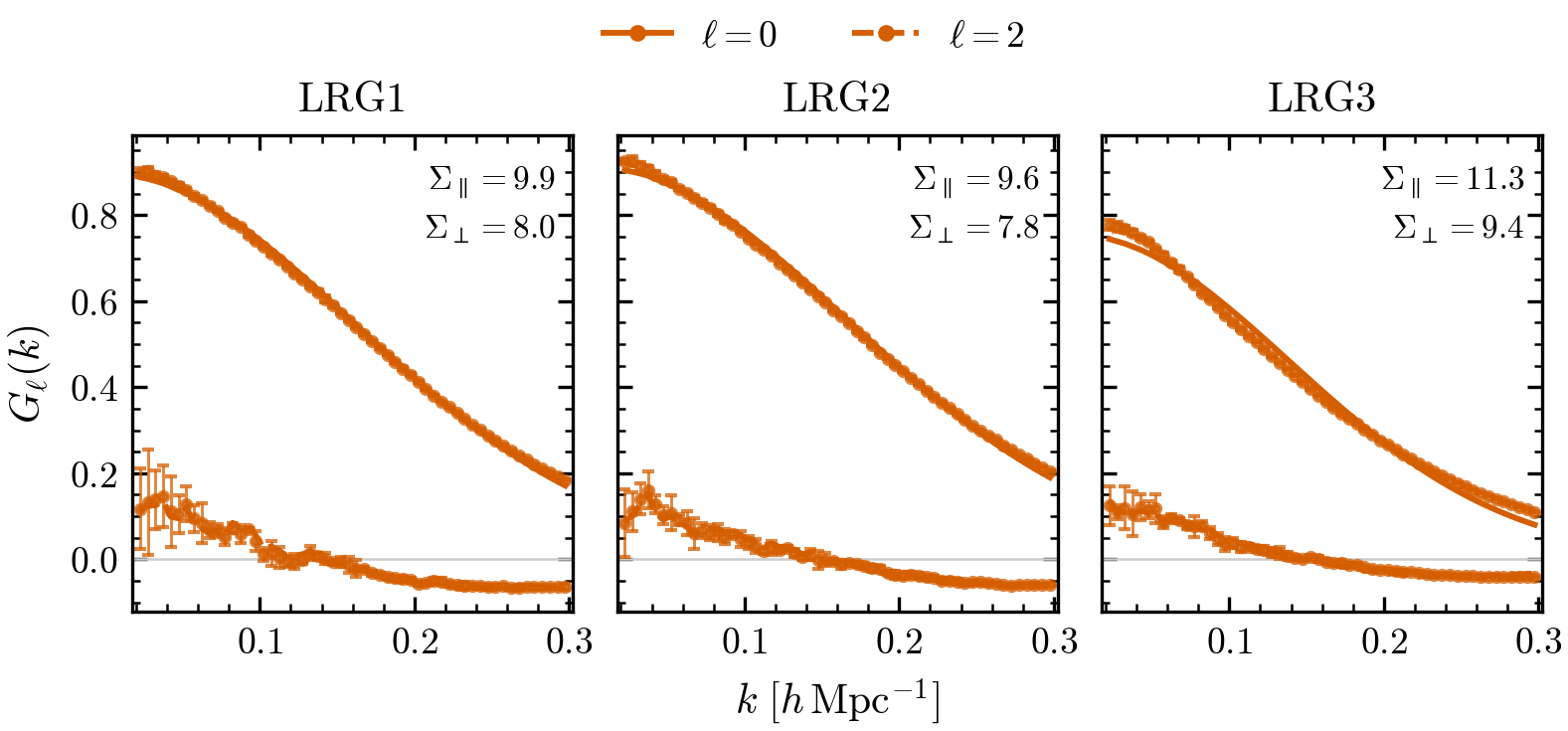}
\caption{Propagator calibration for the LiFT model priors across all three LRG bins. Points show the Legendre multipoles $G_\ell(k) = \frac{2\ell+1}{2}\int_{-1}^{1}{\rm d}\mu\, G(k,\mu)\mathcal{L}_\ell(\mu)$ of the measured propagator $G(k,\mu)$ (\cref{eq:LiFT-propagator}) for $\ell=0,2$, averaged over the training-seed mocks. Seed-to-seed scatter is shown as error bars, while lines show the best-fit model.}
\label{fig:propagator-fits}
\end{figure}

\Cref{fig:propagator-fits} shows the measured propagator and its fit together with the fitted damping scales in each bin; the fitted response amplitudes are $(b_0, b_2) \simeq (0.84,\,0.17)$, $(0.86,\,0.17)$, and $(0.69,\,0.19)$ in LRG1, LRG2, and LRG3, respectively. We do note that LiFT's calibrated damping appears to exceed the standard reconstruction values. This is again because LiFT is trained under an MSE loss, so its optimal solution is the mean of the linear fields consistent with the observed data; on scales where the galaxy field is shot-noise dominated this simply reverts to the prior mean, which is zero. The network therefore shrinks poorly constrained modes toward zero, and as a result its response rolls off considerably faster than the transfer function of standard reconstruction. This larger damping consequently does not indicate additional smearing of the acoustic feature: the propagator of an MSE-optimal estimator satisfies $G = r^2$, so the roll-off of $G$ simply tracks the loss of correlation with the linear field, a relation we verify for LiFT on the held-out simulations (\cref{fig:wiener_relation}). Then, as shown in \cite{parker2025initial}, the reconstruction response cancels from the BAO Fisher information, which only depends on the cross-correlation coefficient $r(k)$ of \cref{fig:reconstruction-correlation}.

For the prior widths, we fix them to $(\sigma_{b_0}, \sigma_{b_2}, \sigma_{\Sigma_\parallel}, \sigma_{\Sigma_\perp}) = (0.2,\,0.2,\,2.0,\,1.0)$. Here, the damping widths are chosen to match the DESI DR1 standard-reconstruction priors, and all widths exceed the seed-to-seed scatter of the calibration. We apply the same prior centers and widths across all tests (on both held-out HalfDome seeds and the AbacusSummit mocks). We also use these priors when fitting the real DESI DR1 LRG data. We evaluate the sensitivity of the DESI constraints to replacing all four calibrated Gaussian priors with uniform priors in \cref{app:prior-sensitivity}, and find no material difference in central values or errors.

\subsection{Survey geometry}
\label{sec:window}

The measured multipoles are modulated by the survey geometry \citep{wilson2017rapid, beutler2021unified} according to
\begin{equation}
    P^{\rm con}_\ell(k)
    = \sum_{\ell'} \int \mathrm{d}q\, q^2\, W_{\ell\ell'}(k,q)\, P_{\ell'}(q),
\end{equation}
where $q$ represents an integration variable over $k$. For LiFT, its fields are defined uniformly on the mesh. Therefore, the survey geometry is just the binary window $W(\mathbf{x})$. This survey geometry enters our analysis twice: (1) it couples Fourier modes in the covariance (\cref{sec:cov}), and (2) it must be convolved with the model when fitting (\cref{sec:LiFT-model}). For the covariance, we use the native geometry computation within \textsc{TheCov}\footnote{https://github.com/cosmodesi/thecov}, and place synthetic unit-weight points per voxel inside the mask, creating a stratified Monte Carlo sampling of the mask. We note that the noise of this sampling is negligible at the $64^3$ resolution of the window kernels. For the window matrix, we use the multi-scale estimation method from \cite{beutler2021unified} and compute the window directly from the grid using \textsc{pypower}. 

\subsection{Covariance matrix}
\label{sec:cov}
We evaluate the covariance of the reconstructed power spectra analytically. Because the power spectrum is quadratic in the density field, its covariance depends on the four-point function, which separates into a disconnected and a connected component, where
\begin{align}
    \operatorname{Cov}^{\rm disc}
    \left[
        \widehat P(\mathbf{k}),\widehat P(\mathbf{k}')
    \right]
    ={}&
    \left|
        \left\langle
            \delta(\mathbf{k})\delta(-\mathbf{k}')
        \right\rangle
    \right|^2
    +
    \left(\mathbf{k}'\leftrightarrow-\mathbf{k}'\right),
    \nonumber\\
    \operatorname{Cov}^{\rm conn}
    \left[
        \widehat P(\mathbf{k}),\widehat P(\mathbf{k}')
    \right]
    ={}&
    \left\langle
        \delta(\mathbf{k})\delta(-\mathbf{k})
        \delta(\mathbf{k}')\delta(-\mathbf{k}')
    \right\rangle_{\rm c}.
    \label{eq:connected_disconnected_covariance}
\end{align}
Here, the connected four-point function is the trispectrum, which captures the non-Gaussian information in the field; if the field is purely Gaussian, the trispectrum vanishes.

LiFT explicitly targets the Gaussian linear field, and its output is well approximated as Gaussian. Indeed, in \cite{bayer2026field} it was shown that the numerical covariance measured from 900 independent realizations of the neural-network reconstruction matches the analytic disconnected covariance to a percent level over the fitted BAO scales. We therefore only retain the disconnected contribution; the near-Gaussian one-point statistics of the LiFT field on the data (\cref{app:shot-noise}) are consistent with this treatment. For standard reconstruction in the results below, we follow the DESI Fourier-space convention and use only the disconnected term to approximate the covariance, which also mirrors our handling of LiFT. This treatment is approximate and may underestimate the errors of standard reconstruction. 

The survey geometry of \cref{sec:window} introduces off-diagonal terms even when considering only the disconnected term in the covariance, since $W(\mathbf{x})$ couples different Fourier modes by convolving the power spectrum with the window. We compute the windowed covariance for LiFT with \textsc{TheCov}, which implements the analytic frameworks of \cite{wadekar2020galaxy,2019JCAP...01..016L}. Specifically, assuming the power spectrum only varies slowly over the Fourier-space width of the window, the disconnected covariance of the multipoles is expressed as 
\begin{equation}
    \operatorname{Cov}^{\rm disc}_{\ell\ell'}(k_i,k_j) = \sum_{L,L'}\mathcal{Q}^{LL'}_{\ell\ell'}(k_i,k_j)P_L(k_i)P_{L'}(k_j),
    \label{eq:thecov_windowed_covariance}
\end{equation}
where $\ell,\ell'\in\{0,2\}$ are the fitted multipoles, the sums run over $L,L'\in\{0,2,4\}$, and $\mathcal{Q}^{LL'}_{\ell\ell'}$ contains the window factors. These in turn are built from the Fourier transform of $W^2$.

Because the shot noise of a survey has known Poisson properties that are non-Gaussian, one typically subtracts out shot-noise from $P_L$, and then adds additional terms in \cref{eq:thecov_windowed_covariance} in order to model the shot noise separately. However, LiFT predicts a continuous field which carries no Poisson shot noise, since it acts as a Wiener filter for Gaussian fields (\cref{app:mse}), and therefore inverse-variance-weights the data, leading to suppression of power in the high-$k$ noise-dominated regime. As a result, we do not subtract out shot noise from $P_L$, and do not include the separate shot-noise terms for LiFT; any residual noise in the reconstruction instead enters \cref{eq:thecov_windowed_covariance} through the measured $P_L$\footnote{In practice we remove the shot-noise terms by setting $\alpha=-1$ in \textsc{TheCov}, which nulls their $(1+\alpha)$ and $(1+\alpha)^2$ prefactors.}. We verify this premise directly on the DESI DR1 data in \cref{app:shot-noise}, where we show that the excess kurtosis of the LiFT field is three orders of magnitude below standard reconstruction and a corresponding Poisson null test. For standard reconstruction, we do keep these additional terms, and compute the geometry from the weighted random catalogs, generally following the DESI convention to compute the covariance matrix \cite{forero2025analytical}. 

\subsection{Fitting the data}
\label{sec:fitting}

We combine the measured multipoles, window-convolved model, calibrated priors, and analytic covariance in a Gaussian likelihood implemented in \textsc{desilike}\footnote{https://github.com/cosmodesi/desilike}, the framework used for the official DESI DR1 analysis \citep{adame2025desi}. We fit on each LRG bin independently by fitting the window-convolved model to the measured monopole and quadrupole of the data with a Gaussian likelihood, where the covariance is defined above. We perform the fit over $0.02 < k < 0.30\,h\,{\rm Mpc}^{-1}$ in bins of width $\Delta k = 0.005\,h\,{\rm Mpc}^{-1}$, matching the DESI DR1 baseline. As in DESI DR1, the additive spline broadband amplitudes are not sampled but solved analytically at each likelihood evaluation. We run \textsc{emcee} \citep{foreman2013emcee}, as implemented 
in \textsc{desilike}, with eight independent chains of 40 walkers. We discard the first half of each chain as burn-in. We quote $(\alpha_{\rm iso}, \alpha_{\rm AP})$ marginalized over all remaining parameters.

\section{Evaluation}
\label{sec:eval}

We evaluate LiFT extensively before applying it to the DESI DR1 LRGs. First, we evaluate the method on held-out HalfDome simulations (\cref{sec:held-out}), which were not seen during training but are drawn from the same distribution. Next, we evaluate on two suites of AbacusSummit mocks \cite{maksimova2021abacussummit}. These mocks are significantly different than the training simulations on which LiFT was trained, as AbacusSummit differs from HalfDome in its gravity solver, halo finder, HOD choices, and cosmology. We test on two suites from AbacusSummit: (1) a modified version of the complete mocks (\cref{sec:abacus-complete}), which tests robustness to misspecification of these effects; and (2) the altMTL mocks \cite{lasker2025production} (\cref{sec:abacus-altmtl}), which test our simplified treatment of fiber assignment against the full DESI fiber-assignment history forward modeled with altMTL. Finally, on these same AbacusSummit altMTL mocks, we test the network's robustness to a distorted distance-redshift mapping in \cref{sec:abacus-warped}. In all cases, we process each mock through the identical pipeline used for the real data (\cref{sec:bao}), and compare against standard reconstruction fit with the DESI DR1 BAO model \cite{adame2025desi}. We note that the improvements shown in this section using LiFT may be an underestimate of the actual improvements because we have assumed a disconnected covariance matrix approximation of standard reconstruction (see \cref{sec:cov} for a detailed discussion). 

\begin{figure}[t]
    \centering
    \includegraphics[width=\textwidth]{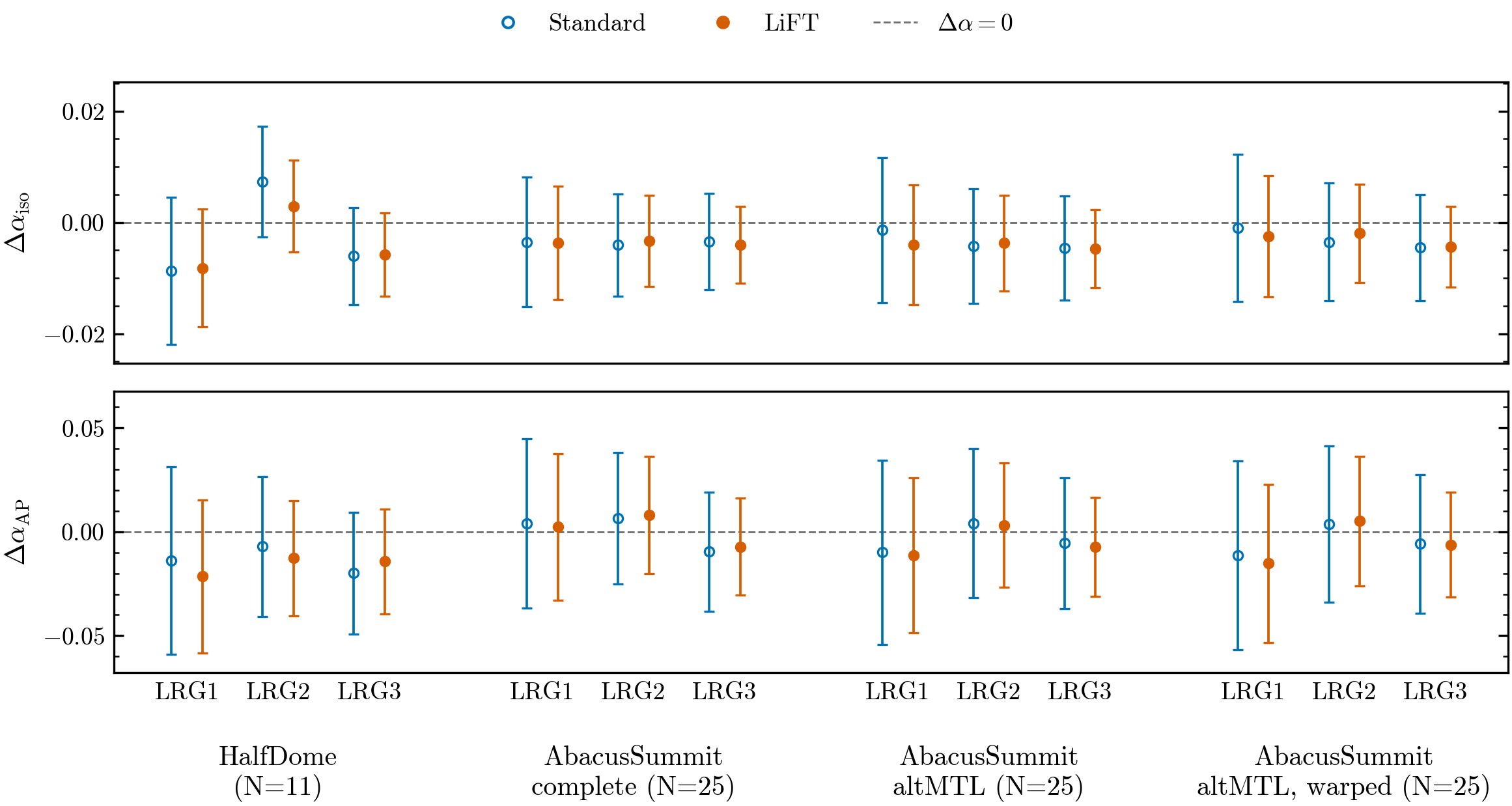}
    \caption{Recovered shifts in $\alpha_{\rm iso}$ and $\alpha_{\rm AP}$, for standard reconstruction (blue, open) and LiFT (orange, filled), in each of the three LRG redshift bins. Results are shown for the held-out HalfDome simulations (left), the complete AbacusSummit mocks (center), and the altMTL AbacusSummit mocks (right). Additionally, we re-analyze the altMTL AbacusSummit mocks with a distorted distance-redshift mapping, with $\Delta\alpha$ measured relative to the injected values of \cref{eq:alpha-expected}. Here, the points are the mean center and error bars are the mean per-realization posterior widths. Across all tests, LiFT gives consistently tighter constraints.}
    \label{fig:whisker-qiso-qap}
\end{figure}

\subsection{HalfDome Simulations}
\label{sec:held-out}

We first evaluate LiFT on the $11$ HalfDome lightcones using the $3$-fold cross-validation networks described above in \cref{sec:training}. For this test, we populate each realization with a fixed HOD chosen from the accepted ensemble as the one whose predicted $P_0$ and $P_2$ most closely match the DESI DR1 LRG measurements. Using the $3$-fold cross-validation scheme ensures that each seed is fit by a network that has never seen it during training or prior calibration. \Cref{fig:whisker-qiso-qap} (left) and \cref{tab:evaluation} provide the full numerical results on the recovered dilation parameters. We find that LiFT improves constraints on $\alpha_{\rm iso}$ by $20\%$, $17\%$, $15\%$ and on $\alpha_{\rm ap}$ by $18\%$, $17\%$, $14\%$ over standard reconstruction across the three LRG bins. These improved mean posterior uncertainties are also roughly reflected in seed-to-seed scatter of the recovered parameters, as the empirical scatter of LiFT is smaller than that of standard reconstruction overall.

\subsection{AbacusSummit}
\label{sec:abacus}
The $25$ AbacusSummit mocks provide a stringent out-of-distribution test of LiFT, as essentially every ingredient of the AbacusSummit forward model differs from that of the HalfDome training simulations:
\begin{itemize}
    \item Gravitational evolution is solved using a high resolution $N$-body code, rather than with FastPM's low resolution particle-mesh scheme.
    \item Halos are identified with the CompaSO spherical-overdensity finder \cite{hadzhiyska2022compaso} rather than with HalfDome's RFOF halo finder.
    \item Galaxies are assigned via a fixed HOD calibrated independently on DESI data \cite{yuan2024desi}.
    \item Planck 2018 cosmology \cite{aghanim2020planck} is used instead of the Planck 2015 \cite{ade2016planck} used in HalfDome.
    \item For the altMTL setting, the mocks include a forward model of the DESI fiber-assignment process rather than uniform statistical thinning.
\end{itemize}
In order for LiFT to maintain unbiased recovery on these mocks, the network must be able to transfer across all of these effects. We evaluate LiFT on two different settings of the AbacusSummit mocks: complete, which does not forward model fiber assignments, and altMTL, which does forward model fiber assignments. This is in order to separate the effects of our simplified treatment of fiber assignment in our forward model (\cref{sec:data:selection}). We note that we modify the complete mocks to match shot noise of HalfDome and altMTL, as described below.

\subsubsection{Complete Mocks}
\label{sec:abacus-complete}
The complete mocks carry the DR1 footprint and radial selection, but no fiber assignment. Because DR1 assigns fibers to only ${\sim}69\%$ of LRG targets, the native complete mocks from the AbacusSummit suite have much higher number density, and thus much lower shot noise, than the DESI DR1 sample and than their altMTL equivalents. To account for this, we follow the thinning used for HalfDome (see \cref{sec:data:selection}) to match the number density of the complete mocks to their phase-paired altMTL equivalent. Thus, the resulting mocks in this test have the same angular coverage and number density as the altMTL mocks, but lack the full fiber assignment history.

\Cref{fig:whisker-qiso-qap} (center) and \cref{tab:evaluation} show the recovered dilation parameters. Despite the many differences noted above, LiFT remains as unbiased as standard reconstruction. Moreover, the precision gains persist: LiFT improves constraints on $\alpha_{\rm iso}$ by $13\%$, $11\%$, $21\%$ and on $\alpha_{\rm AP}$ by $13\%$, $11\%$, $19\%$ across the three bins. As a result, we conclude that LiFT transfers well across N-body, halo finder, HOD, and cosmology without developing biases or losing its more precise constraining power.  

\subsubsection{altMTL Mocks}
\label{sec:abacus-altmtl}
The altMTL mocks add DR1 tile ordering, target priorities, and fiber collisions to the complete mocks above using the altMTL framework \cite{lasker2025production}. As a result, incompleteness is density-dependent, as galaxies in overdense regions compete for fibers. We note that for this test and in general, LiFT has never been trained in a setting with realistic fiber assignments, and as a result has never seen density-dependent completeness variations. 

\Cref{fig:whisker-qiso-qap} (right) and \cref{tab:evaluation} show the recovered dilation parameters. Despite the out-of-distribution realistic fiber assignment, LiFT remains as unbiased as standard reconstruction. Moreover, as above, the precision gains persist: LiFT improves constraints on $\alpha_{\rm iso}$ by $17\%$, $17\%$, $25\%$ and on $\alpha_{\rm AP}$ by $16\%$, $17\%$, $24\%$ across the three bins. We conclude that our simplified treatment of fiber assignment neither biases LiFT nor erodes its constraining-power advantage.

\subsection{Alcock-Paczynski distortion test}
\label{sec:abacus-warped}
We evaluate the network's response to a misspecified cosmology used for converting observed galaxy redshifts and angles into comoving coordinates. Concretely, given the Hubble and transverse comoving distances in a flat $\Lambda$CDM universe,
\begin{equation}
    D_H(z) = \frac{c}{H_0\sqrt{\Omega_{\rm m}(1+z)^3 + 1 - \Omega_{\rm m}}},
    \qquad
    D_M(z) = \int_0^z \mathrm{d}z'\, D_H(z'),
    \label{eq:dh-dm}
\end{equation}
lowering $\Omega_{\rm m}$ increases both, and increases $D_H$ by more than $D_M$ at a given redshift. As a result, radial separations in the wrong frame are rescaled by $D_H^{\rm fid}/D_H^{\rm true}$ and transverse separations by $D_M^{\rm fid}/D_M^{\rm true}$. Therefore, we can write the expected dilations under this transformation as 
\begin{equation}
    \alpha_\parallel^{\rm exp}
    = \frac{D_H^{\rm true}(z_{\rm eff})}{D_H^{\rm fid}(z_{\rm eff})},
    \qquad
    \alpha_\perp^{\rm exp}
    = \frac{D_M^{\rm true}(z_{\rm eff})}{D_M^{\rm fid}(z_{\rm eff})}.
    \label{eq:alpha-expected}
\end{equation}
Here, we choose $\Omega_{\rm m}^{\rm fid} = 0.2344$ (for AbacusSummit, the truth is $\Omega_{\rm m}^{\rm true} = 0.3152$). Other than $\Omega_m$, all other parameters are held at their AbacusSummit values. This corresponds to $(\alpha_{\rm iso}^{\rm exp},\alpha_{\rm AP}^{\rm exp}) = (0.962,\,0.971)$, $(0.950,\,0.962)$, and $(0.940,\,0.956)$ in LRG1, LRG2, and LRG3\footnote{Note as well that the mapping is not volume-preserving, $\mathrm{d}^3x' = \mathrm{d}^3x/\alpha_{\rm iso}^3$, so the number density seen by the network is also lowered by $\alpha_{\rm iso}^3$.}. 

We perform this test on the 25 AbacusSummit altMTL mocks. We apply the mapping to the galaxies and randoms for each mock, and resample the survey mask and coverage map onto the wrong-frame grid by nearest-voxel lookup. We then rerun the full pipeline in the wrong frame, using standard reconstruction with misspecified cosmology and the same LiFT trained on HalfDome simulations. \Cref{fig:whisker-qiso-qap} (right) shows the recovered dilation parameters relative to the injected values. LiFT recovers the injected distortion with the same fidelity as standard reconstruction, and its precision advantage is unchanged.

\subsection{Empirical coverage and bias}

In Table \ref{tab:evaluation} we report 
the posterior means for all the 
mock analyses. We see that the recovered values are 
on average slightly below 1. This bias in the mean is always 
below one standard deviation for one realization, but 
in a few cases 
the statistical significance of the bias on 
the mean of all the realizations is statistically significant. 
This bias is shared by LiFT and standard reconstruction alike, indicating a shared origin rather than a miscalibration specific to the network-based analysis---it is likely caused by the survey geometry and 
incompleteness, as it is significantly reduced for the standard reconstruction in DR2 relative to DR1 \cite{DESI_dr2_val}. We also note that, given the small number of realizations, it is difficult to draw conclusions about these offsets, as individual realizations can appreciably shift the ensemble mean (see Figures \ref{fig:individual-halfdome}--\ref{fig:individual-warped} for the fits to each individual realization). We also report 
results of the AP rescaling test, resulting in the correctly rescaled values of $\alpha$ that are very different from the training values---the results indicate that 
the mean value is recovered close to the true value, with a similar bias to the fiducial 
cosmology case. This confirms that the small bias we observe is due to the survey 
itself, and does not depend on the true values of 
$\alpha$.

In Table \ref{tab:ensemble-mse}
we also report the realization-to-realization scatter of the central values, which enables us to compare the reported posterior widths with the empirical scatter. Across all suites, the empirical scatter is roughly consistent with the mean posterior widths. To quantify this, we pool the six (bin, parameter) combinations of the held-out HalfDome and altMTL suites and compute the $N$-weighted mean of the ratio of the empirical scatter to the mean per-realization posterior width, $\langle\mathrm{std}(\alpha)/\sigma_\alpha\rangle$. We find $1.06$ for LiFT and $1.08$ for standard reconstruction. The expected variance for 216 
realizations is $(2/216)^{1/2} \sim 0.10$, indicating that both estimators' mock uncertainties agree with the mock-to-mock scatter well within their $1\sigma$ intervals. 

We note that there is some duplication of survey volume in our mocks due to the finite size of simulation boxes relative to survey volume of DR1: in HalfDome, $9.5\%$
of the LRG3 volume is replicated (and less in LRG1 and LRG2); in
AbacusSummit, which uses a smaller box, the replicated fractions are $17.3\%$,
$12.5\%$, and $26.4\%$ in LRG1, LRG2, and LRG3 respectively\footnote{We compute the replicated volume by folding the survey back into a single box period and counting, at
each position within the box, how many survey cells map onto it.}. Typically this will lead to an increase of empirical scatter relative to the posterior: there are fewer independent realizations of BAOs than the survey volume predicts. Our coverage tests suggest this is not a large effect, but it is possible that the slight excess of empirical scatter ($1.06$ and $1.08$) is caused by this. We emphasize that this is not a concern for the 
analysis of the actual DR1 DESI survey, which does not suffer from this issue. 
For a more thorough coverage test, we refer the reader to \cite{bayer2026field}, which performed coverage tests over $900$ mock realizations of this methodology, showing 
good coverage. 

\section{Results on DESI DR1 LRGs}
\label{sec:desi_results}

We apply LiFT to the DESI DR1 LRGs. We use the network trained on all eleven HalfDome seeds, which is the identical model validated on the out-of-distribution AbacusSummit mocks above (\cref{sec:abacus}). We first present the reconstructed clustering and BAO fits, then report the resulting BAO constraints and distance measurements, and finally compare these measurements with previous analyses.

\subsection{Reconstructed clustering and BAO fits}
\label{sec:desi_clustering}

Following the inference scheme in \cref{sec:inference}, we feed LiFT subsets of the observed DESI galaxy overdensity field on a meshed grid, along with its RecSym standard reconstruction (performed on the full LRG sample) and the additional survey context channels described in \cref{sec:architecture}. We then tile LiFT's predictions across the analysis grid to obtain an estimate of the $z=0$ linear density field over the full DR1 LRG volume. 

\begin{figure}[t]
    \centering
    \includegraphics[width=\textwidth]{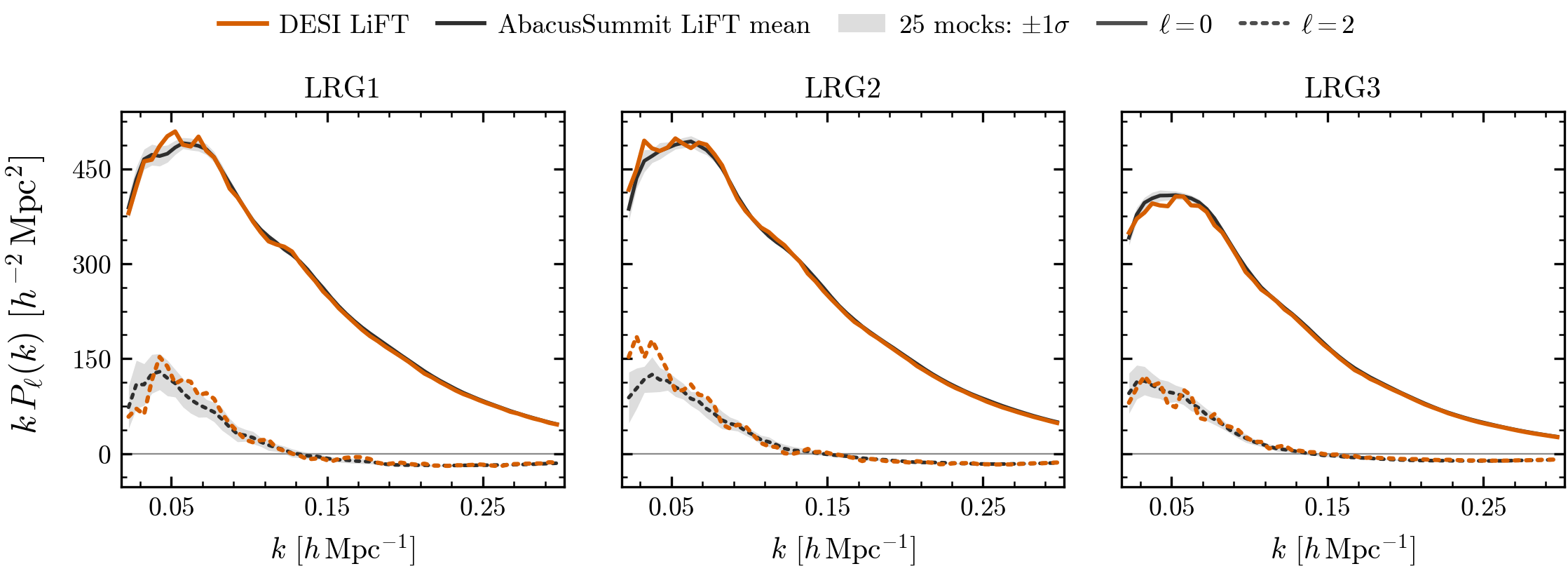}
    \caption{Power spectrum multipoles of the LiFT reconstruction of the DESI DR1 LRG data (orange), compared with the mean (black) and $\pm1\sigma$ spread (gray bands) over the 25 altMTL AbacusSummit mocks processed with the identical pipeline.}
    \label{fig:desi-vs-mocks}
\end{figure}

From this field, we measure the power spectrum multipoles in each LRG bin (\cref{sec:pk}) and plot them in \cref{fig:desi-vs-mocks}. Ultimately, the LiFT multipoles display the expected features of an estimate of the real-space linear field. Specifically, the LiFT quadrupole is strongly suppressed and decays rapidly with $k$, due to LiFT removing redshift-space distortions; there is still some residual quadrupole, reflecting the line-of-sight anisotropy of the network transfer. Moreover, the LiFT monopole is significantly lower than that of the standard-reconstructed galaxy field. This reflects LiFT's target of a real-space linear density field normalized to $z=0$, which has a different normalization from the biased galaxy field in redshift space. Additionally, the LiFT monopole fall-off at high $k$ is not only a constant bias shift, but also reflects the Wiener-filter-like roll-off of the network response on shot-noise-dominated scales (\cref{sec:priors}). \Cref{fig:desi-vs-mocks} also provides, for comparison, the measured multipoles of LiFT applied to the AbacusSummit altMTL mocks; we note that LiFT's reconstructed DESI multipoles show broadly similar monopole shapes and quadrupole suppression to the mocks. 

Once measured, we fit the multipoles with the model described in \cref{sec:LiFT-model}. As in the held-out evaluations on the HalfDome and AbacusSummit suites, we use the priors on $(b_0, b_2, \Sigma_{\parallel}, \Sigma_{\perp})$ calibrated on the HalfDome training simulations (\cref{sec:priors}), and the analytic covariance is computed using the LiFT reconstructed multipoles as described in \cref{sec:cov}. The model describes the data well, with $\chi^2/{\rm ndof} = 96.0/90$, $71.6/90$, and $101.9/90$ in LRG1, LRG2, and LRG3, respectively.

\begin{figure}[t]
\centering
\includegraphics[width=\textwidth]{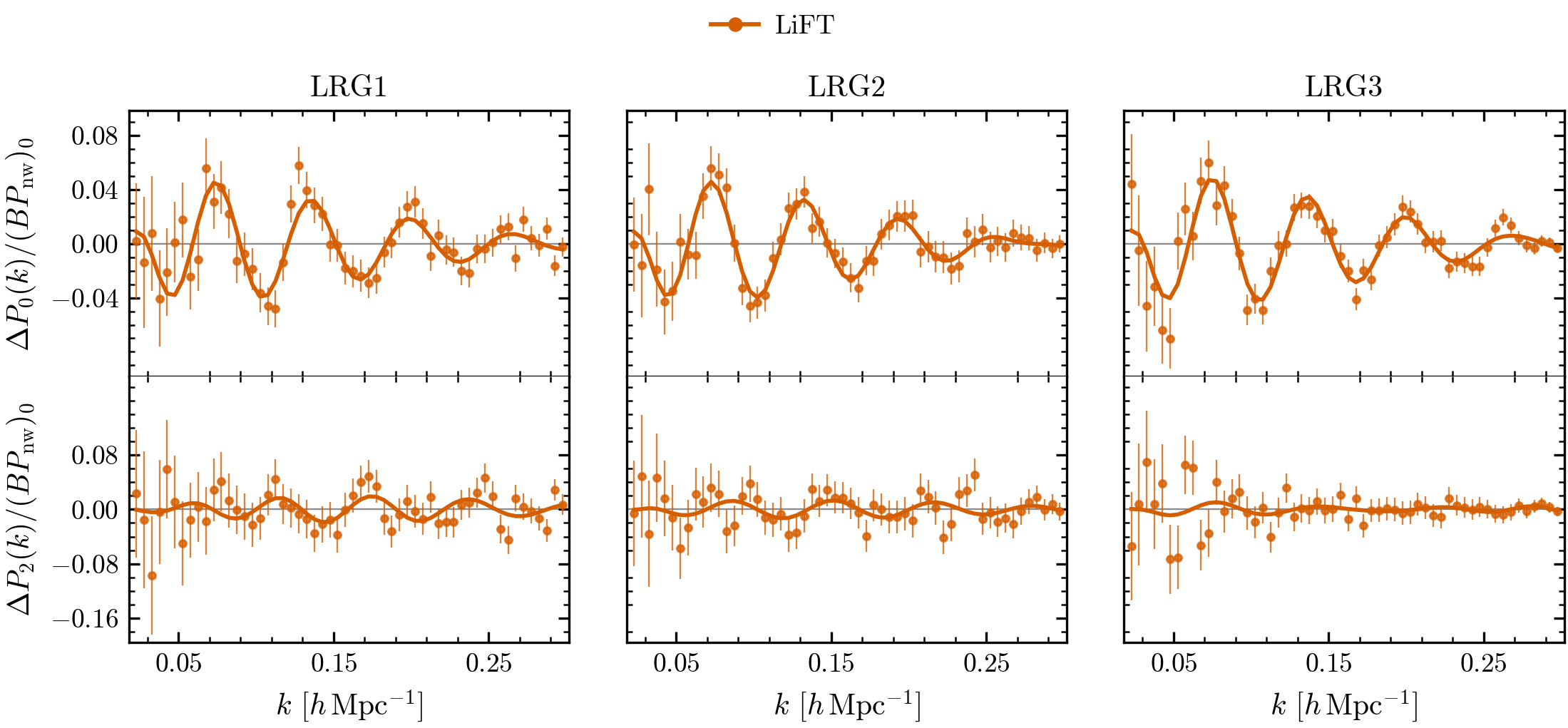}
\caption{Isolated BAO signature in the LiFT monopole of the DESI DR1 LRG data. Points show $\Delta P_\ell = P_{\ell} - (BP_{\rm nw})_\ell - D_\ell$, normalized by $(BP_{\rm nw})_0$.}
\label{fig:desi_wiggles}
\end{figure}

In addition to the overall multipole fit, we isolate the acoustic signature in the LiFT multipole by subtracting the smooth component of the model evaluated at the posterior-mean parameters from the measured multipoles. Specifically, we let $\Delta P_\ell = P_{\ell} - (BP_{\rm nw})_\ell - D_\ell$, and normalize this quantity by the smooth monopole template $(BP_{\rm nw})_0$. \Cref{fig:desi_wiggles} shows the isolated acoustic signature, with clear oscillations in the monopole and a much weaker acoustic component in the quadrupole. Additionally, we find that LiFT achieves significance of the BAO feature of $7.4\sigma$, $8.1\sigma$, and $10.2\sigma$ in LRG1, LRG2, and LRG3, respectively, compared to $6.4\sigma$, $6.8\sigma$, and $8.7\sigma$ for standard reconstruction in those three bins (cf. figure 11 of \cite{adame2025desi}); for this diagnostic, we fit the monopole alone, fixing $\alpha_{\rm AP}=1$, fixing $b_2$ to its calibrated prior mean, and setting the unfitted multipoles’ broadband coefficients to zero, after which we quote $\sqrt{\Delta\chi^2}$ from the data $\chi^2$ values at the independently optimized BAO and no-BAO posterior modes.

\subsection{BAO constraints}
\label{sec:des_bao_constraints}

We compare the fitted $(\alpha_{\rm iso}, \alpha_{\rm AP})$ parameters from the LiFT reconstruction with the DESI DR1 standard reconstruction primary results. Throughout this section, we compare against the DESI DR1 configuration-space ($\xi(s)$) results, rather than its Fourier-space counterpart. This choice of baseline is motivated by the fact that DESI's Fourier-space analytic covariance did not reach the desired level of agreement with the mock-based covariances \citep{adame2025desi,forero2025analytical}. Thus, both DR1 and DR2 use configuration-space as their headline results \cite{adame2025desi, Abdul_Karim_2025}, and calibrate their covariance with a jackknife-calibrated shot-noise rescaling that acts as a proxy for the missing connected (trispectrum) contribution \cite{rashkovetskyi2025semi}. This concern, and the need for additional calibration, does not carry over to our Fourier-space analysis of LiFT: because LiFT produces a reconstruction of the Gaussian linear field, its covariance is well-approximated by the analytic disconnected covariance alone, which we have validated at the percent level against $900$ independent realizations of the reconstruction in our previous analyses \cite{bayer2026field} and whose premise we check on the DR1 data in \cref{app:shot-noise}. For a more complete discussion, see \cref{sec:cov}. We note nonetheless that our standard-reconstruction pipeline applied in Fourier space reproduces the DESI DR1 Fourier-space measurements (\cref{app:desi-pk-validation}).

We provide the quantitative results of these fits in \cref{tab:desi_alpha_comparison}. LiFT improves the constraints on $\alpha_{\rm iso}$ by $13\%$, $24\%$, and $30\%$, and on $\alpha_{\rm AP}$ by $5\%$, $22\%$, and $30\%$, across the three LRG bins, with the corresponding error-product Figure of Merit improvements of 1.2, 1.7 and 2.0. Given that $\sigma_\alpha \propto V_{\rm eff}^{-1/2}$, matching this reduction in the product of errors by increasing survey volume alone would require the same factors in volume at fixed methodology. These gains are broadly comparable to those found in the held-out mock suites. We note that the improvement is notably smaller in LRG1, particularly for $\alpha_{\rm AP}$. This variation may reflect both differences in reconstruction efficiency across the samples and statistical fluctuations in the measured BAO feature.

In addition to the tightened precision, we find that the recovered central values are consistent between standard reconstruction and LiFT: the shifts are $\Delta\alpha_{\rm iso} = +0.008$, $+0.001$, and $-0.008$, and $\Delta\alpha_{\rm AP} = +0.028$, $-0.012$, and $+0.003$, in the LRG1-3 bins respectively. For reference, on the altMTL mocks, we find a root mean scatter of $0.010$, $0.007$, and $0.007$ in $\alpha_{\rm iso}$ and $0.028$, $0.021$, and $0.015$ in $\alpha_{\rm AP}$ for $\alpha_{\rm LiFT}-\alpha_{\rm std}$: the observed shifts are at most $1.1$ times the expected statistical scatter between the two estimators. 

\begin{table}
\centering
\caption{Mean values and standard deviations of the BAO scaling parameters, $\alpha_{\rm iso}$ and $\alpha_{\rm ap}$, from fits of the field-level LiFT $P(k)$ analysis of the DESI DR1 LRG data, compared with the DESI DR1 standard reconstruction $\xi(s)$ results (table 15 of \cite{adame2025desi}). All uncertainties are statistical only. We also show the fractional reduction of the LiFT uncertainties relative to the standard reconstruction values and the overall improvement in Figure of Merit (FoM) defined as the ratio of the product of $\alpha_{\rm iso}$ and $\alpha_{\rm ap}$ errors between standard reconstruction and LiFT.}
\label{tab:desi_alpha_comparison}
\setlength{\tabcolsep}{4pt}
\begin{tabular}{lcccccc}
\toprule
Bin & Method & $\alpha_{\rm iso}$ & $\alpha_{\rm AP}$ & $\chi^2/{\rm dof}$ & $(\Delta\sigma_{\rm iso},\,\Delta\sigma_{\rm AP})$ & FoM \\
\midrule
\multirow{2}{*}{LRG1} &  Standard & $0.9792\pm0.0113$ & $0.9148\pm0.0372$ & $41.8/39$ & --- & ---\\
& LiFT & $0.9872\pm0.0099$ & $0.9432\pm0.0355$  & $96.0/90$ & $(13\%,\,5\%)$ & 1.2\\
\addlinespace
\multirow{2}{*}{LRG2} & Standard & $0.9661\pm0.0117$ & $1.0459\pm0.0432$ & $46.5/39$ & --- &---\\
&  LiFT & $0.9671\pm0.0089$ & $1.0342\pm0.0339$ & $71.6/90$ & $(24\%,\,22\%)$ & 1.7\\
\addlinespace
\multirow{2}{*}{LRG3} &  Standard & $1.0042\pm0.0090$ & $1.0027\pm0.0299$ & $46.2/39$ & --- &---\\
&  LiFT & $0.9961\pm0.0063$ & $1.0053\pm0.0210$ &  $101.9/90$ & $(30\%,\,30\%)$ & 2.0\\
\bottomrule
\end{tabular}
\end{table}

\begin{figure}
\centering
\includegraphics[width=\textwidth]{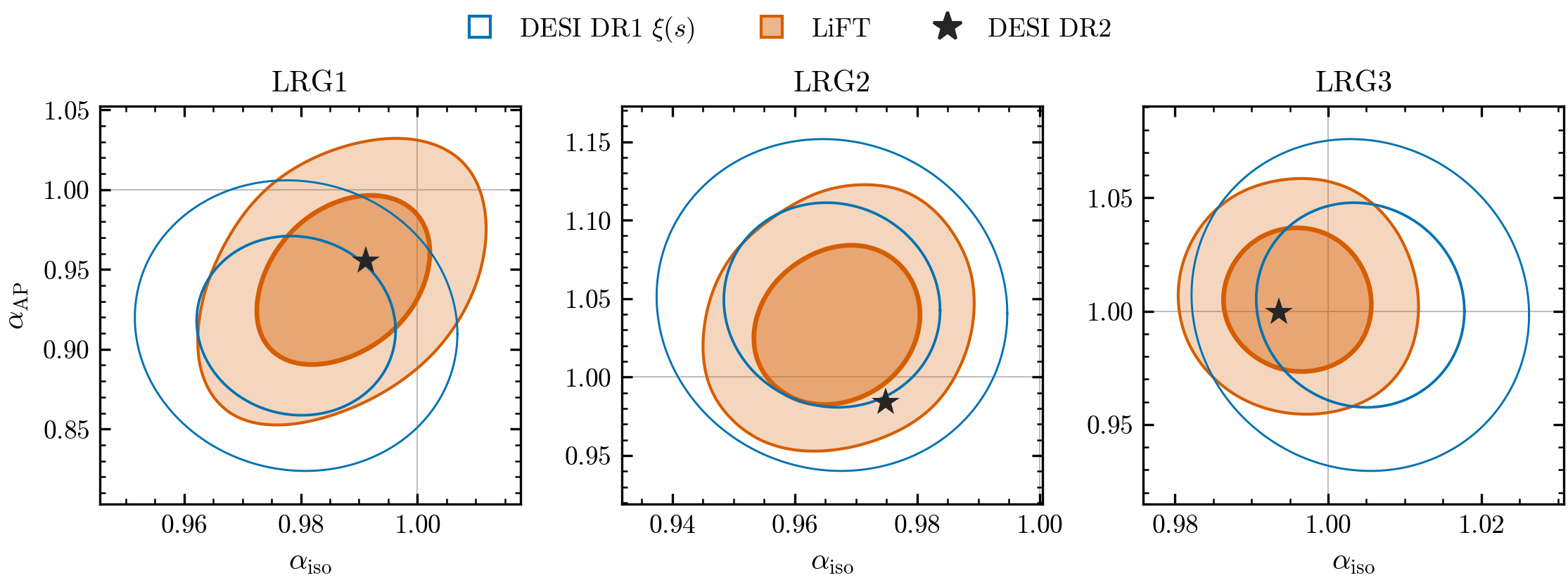}
\caption{68\% and 95\% credible regions for the LiFT fits of the DESI DR1 LRG data (orange, filled), compared with the DESI DR1 $\xi(s)$ results (blue, open). The $\xi(s)$ contours are Gaussian approximations constructed from the published posterior means, standard deviations, and correlation coefficients of table 15 of \cite{adame2025desi}. Stars mark the DESI DR2 central values.}
\label{fig:desi_contours}
\end{figure}

\Cref{fig:desi_contours} shows the corresponding 68\% and 95\% credible regions for the LiFT and standard reconstruction fits on the DESI DR1 data for each LRG bin. For reference, we also include the central value of the DESI DR2 measurements for each of the LRG bins \cite{Abdul_Karim_2025}. The DR2 central values are consistent with the LiFT measurements in all three bins. We note that while DR1 is a subset of DR2 and thus the measurements are strongly correlated, it is nevertheless reassuring that the higher-precision LiFT measurements remain compatible with the subsequent more constraining DESI DR2 data. Indeed, the shifts of LiFT central value relative to DR1 standard reconstruction central value are all in the direction of DR2. Despite our improvements, DR2 errors are still considerably smaller than DR1 LiFT, a consequence of a significant increase in volume and number of LRG galaxies between DR1 and DR2. 

Finally, we transform the posterior samples to $(\alpha_\perp,\alpha_\parallel)$, and find correlation coefficients between these parameters of $-0.437$, $-0.421$, and $-0.401$ in LRG1–3, respectively. These roughly match the idealized isotropic BAO Fisher expectation of $-1/\sqrt{6}\simeq-0.408$ \cite{seo2007improved, adame2025desi}, indicating that the recovered constraints exhibit the characteristic transverse-radial BAO degeneracy.

\section{Conclusions}
\label{sec:conclusions}

We have introduced Linear Field Transformer (LiFT), a hybrid method that uses a survey-geometry-aware 3D vision transformer on subgrids of the survey volume to learn non-linear corrections to standard reconstruction, thereby reconstructing a linearized density field from an observed galaxy survey. We construct a suite of HalfDome lightcones that reproduce the DESI DR1 footprint, radial selection, and mean fiber-assignment incompleteness, along with varying HOD parameters, and train LiFT to predict the linear fields of this suite. Once trained, we develop the methodology to enable LiFT's reconstructed linear field to be used to perform BAO analysis; because LiFT targets the near-Gaussian linear field, it admits a simple BAO pipeline, consisting of a simulation-calibrated transfer function and a purely analytic disconnected covariance, while otherwise retaining the DESI DR1 fitting infrastructure.

Before applying LiFT to real data, we demonstrate that LiFT's BAO pipeline produces consistently unbiased constraints and improved precision on both held-out simulations from the same distribution as the training suite, as well as on AbacusSummit mocks, which differ from this training suite in terms of gravity solver, halo finder, HOD, cosmology, and, in their altMTL variant, fiber-assignment history. Moreover, we verify that LiFT correctly recovers the Alcock-Paczynski distortion induced by analyzing the mocks with an incorrect distance-redshift relation.

Once validated, we apply LiFT to the DESI DR1 LRG sample. We find that LiFT improves constraints on $\alpha_{\rm iso}$ by $13\%$, $24\%$, and $30\%$ and on $\alpha_{\rm AP}$ by $5\%$, $22\%$, and $30\%$ over the DESI DR1 $\xi(s)$ results, while the central values remain consistent with both DR1 and DR2 standard reconstruction analyses. Ultimately, the improved Figure of Merit corresponds to a ${\sim}1.2$-$2\times$ increase in effective survey volume at fixed methodology. To our knowledge, this constitutes the first application of neural field-level BAO reconstruction to real spectroscopic survey data.

For future work, the most direct extension will be to extend the analysis to DESI DR2 and DR3, as well as to other tracers such as BGS and ELG galaxies. Provided the precision gains hold, this will enable tighter distance measurements, and subsequently tighter dark-energy constraints. Beyond BAO, LiFT's estimation of the linear density field can also be used as a natural input for a range of other cosmological analyses, including possible full-shape inference, velocity reconstruction for kSZ science \cite{tröster2026velocityformerbrokensymmetrymatchedequivariantgraph}, and constraints on primordial non-Gaussianity \cite{chen2025probingprimordialnongaussianityreconstructing}. We note that the present validation suite only applies so far to BAO, and thus subsequent analyses will require a  dedicated characterization of the network response on the relevant physics; nonetheless, these directions highlight the promise of field-level reconstruction as a general-purpose tool for current and upcoming surveys.

\acknowledgments
We thank Otávio Alves, Xinyi Chen, François Lanusse, Arnaud de Mattia, Michael Rashkovetskyi, Ashley Ross, Alina Sabyr, Natali de Santi, and Hee-Jong Seo for helpful discussions. This research used resources of the National Energy Research Scientific Computing Center (NERSC), a Department of Energy User Facility. LP is supported by the National Science Foundation Graduate Research Fellowship Program (NSF GRFP). This work is supported by NSF CDSE grant number AST-2408026 and NASA TCAN grant number 80NSSC24K0101. 

We acknowledge the use of Claude for assistance with code development and for proofreading the manuscript. All scientific content, analysis choices, and conclusions are the authors' own, and the authors take full responsibility for the contents of this paper.

\bibliographystyle{JHEP}
\bibliography{references,refs}

\newpage

\appendix

\section{HOD Parameters}
\label{app:hod-ensemble}

\Cref{tab:hod-ranges} lists the ranges of the HOD parameters of \cref{eq:hod} spanned by the accepted ensemble described in \cref{sec:data:mocks} across all eleven HalfDome realizations. Each proposal draw for a set of HOD parameters is retained only if it satisfies the number-density, satellite-fraction, and linear-bias selection criteria that are described in \cref{sec:data:mocks}.

\begin{table}
\centering
\caption{%
Ranges of the HOD parameters of \cref{eq:hod} at the three anchor redshifts, taken over all accepted HODs across the eleven HalfDome realizations. Masses are in units of $h^{-1}M_\odot$.}
\label{tab:hod-ranges}
\setlength{\tabcolsep}{5pt}
\begin{tabular}{llccc}
\toprule
Parameter & Description & $z=0.40$ & $z=0.71$ & $z=1.10$ \\
\midrule
$\log M_{\rm cut}$ & Central transition mass       & $[12.76,\,12.95]$ & $[12.73,\,12.88]$ & $[12.47,\,12.84]$ \\
$\sigma_{\log M}$  & Width of central transition   & $[0.12,\,0.17]$   & $[0.18,\,0.23]$   & $[0.25,\,0.30]$   \\
$\log M_0$         & Satellite cutoff mass         & $[13.42,\,13.61]$ & $[13.03,\,13.22]$ & $[12.54,\,12.73]$ \\
$\log M_1$         & Characteristic satellite mass & $[13.69,\,13.99]$ & $[13.55,\,13.77]$ & $[13.27,\,13.68]$ \\
$\alpha$           & Satellite power-law slope     & $[1.13,\,1.29]$   & $[0.97,\,1.13]$   & $[0.77,\,0.93]$   \\
\bottomrule
\end{tabular}
\end{table}

\section{Response of an MSE-optimal field estimator}
\label{app:mse}

Because the LiFT network is trained under an MSE loss, its optimum is the conditional mean of its target. We show below that this gives the reconstruction Wiener-filter-like two-point statistics. Specifically, its propagator satisfies $G=r^2$, so modes that cannot be recovered from the inputs are suppressed toward the prior mean rather than retained as noise.

To see this, consider the contribution of a single cell to the expected training loss (\cref{eq:training-loss}). For any estimator $f$, we can add and subtract the conditional mean, $\bar\delta\equiv\big\langle\delta_{\rm lin}\mid I\big\rangle$, inside the residual to obtain $f-\delta_{\rm lin}=(f-\bar\delta)+(\bar\delta-\delta_{\rm lin})$, where $I=(\delta_{\rm g},\delta_{\rm rec},\bm c)$ is the network input. Expanding the square and taking the expectation therefore yields
\begin{equation}
\big\langle \left[f-\delta_{\rm lin}\right]^2 \big\rangle = \big\langle \left[f-\bar\delta\right]^2 \big\rangle +2\big\langle \left(f-\bar\delta\right)\left(\bar\delta-\delta_{\rm lin}\right)\big\rangle +\big\langle \left[\bar\delta-\delta_{\rm lin}\right]^2 \big\rangle, 
\label{eq:mse_expand}
\end{equation}
where the expectations are taken over the joint distribution of inputs and targets that are given by using the specific mock forward model that we use.

Because both $f$ and $\bar\delta$ are functions of $I$ alone, conditioning on $I$ inside the expectation gives
\begin{equation}
\big\langle(f-\bar\delta)(\bar\delta-\delta_{\rm lin})\big\rangle = \big\langle(f-\bar\delta)\big\langle\bar\delta-\delta_{\rm lin}\mid I\big\rangle\big\rangle =0,
\end{equation}
and so the cross term vanishes. Moreover, the last term of \cref{eq:mse_expand} does not depend on $f$. Since \cref{eq:training-loss} is a non-negatively weighted sum of such per-cell terms, it is minimized by
\begin{equation}
f^\ast \in \underset{f}{\arg\min}\; \big\langle \sum_{\bm x\in\Omega'}w(\bm x)\big[f(I)(\bm x)-\delta_{\rm lin}(\bm x)\big]^2 \big\rangle, \qquad f^\ast(I)(\bm x) = \big\langle\delta_{\rm lin}(\bm x)\mid I\big\rangle, 
\label{eq:fstar}
\end{equation}
which is just the well-known result of the posterior mean of the linear field.

\begin{figure}
    \centering
    \includegraphics[width=0.75\textwidth]{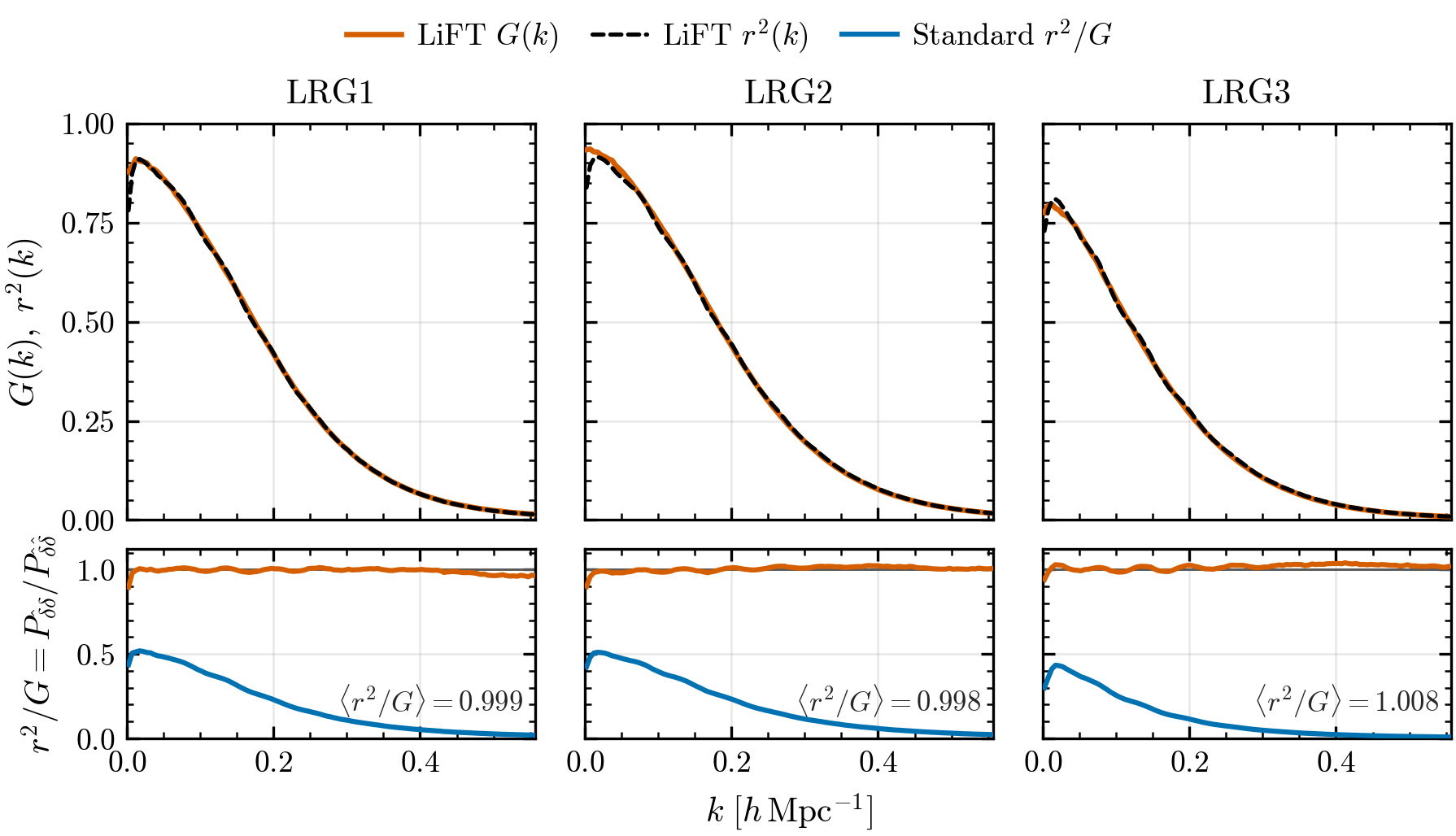}
    \caption{Test of the conditional-mean relation on the held-out cross-validation HalfDome realizations. The upper panels compare the LiFT $G(k)$ with $r^2(k)$ and the lower panels show their ratio. We include standard reconstruction (blue) for reference. }
    \label{fig:wiener_relation}
\end{figure}

The argument above demonstrates the minimum for $f$, however in \cref{eq:mse_expand} we still have an irreducible contribution to the MSE loss which does not depend on $f$; this is the variance of the  conditional-mean residual, $\epsilon\equiv\delta_{\rm lin}-\bar\delta$, since $\langle(\bar\delta-\delta_{\rm lin})^2\rangle=\langle\epsilon^2\rangle$, which is the irreducible information about the linear field not contained in $I$. The same conditioning argument as used above can show that this residual is uncorrelated with the conditional-mean prediction. At the MSE optimum, $\delta_{\rm lin}^{\rm pred}=\bar\delta$, and therefore, for any pair of cells $\bm x$ and $\bm y$,
\begin{equation}
\big\langle \delta_{\rm lin}^{\rm pred}(\bm x)\epsilon(\bm y) \big\rangle = \big\langle \delta_{\rm lin}^{\rm pred}(\bm x) \big\langle\epsilon(\bm y)\mid I\big\rangle \big\rangle =0,
\end{equation}
by definition of $\epsilon$. $\delta_{\rm lin}=\delta_{\rm lin}^{\rm pred}+\epsilon$, 
\begin{equation}
P_{\rm pred,lin} = P_{\rm pred,pred}+P_{\rm pred,\epsilon} = P_{\rm pred,pred}. \label{eq:orth}
\end{equation}
Thus, the auto-power of the conditional-mean prediction equals its cross-power with the true linear field. Substituting $P_{\rm pred,pred}=P_{\rm pred,lin}$ into the definition of $r$ and $G$ gives
\begin{equation}
r^2
=
\frac{P_{\rm pred,lin}^2}
{P_{\rm pred,pred}P_{\rm lin}}
=
\frac{P_{\rm pred,lin}}
{P_{\rm lin}}
=
G.
\label{eq:G_r2}
\end{equation}
We demonstrate that this identity holds for LiFT on the held-out HalfDome cross-validation simulations in \cref{fig:wiener_relation}. Ultimately, we find that $G/r^2 \approx 1$ for LiFT; standard reconstruction by contrast, is not a conditional-mean estimator and therefore $G \neq r^2$ and unrecoverable small-scale power is retained as noise rather than suppressed.

This relationship explains several choices in the main text:
\begin{enumerate}
    \item Modes with $r \to 0$ are shrunk toward the prior mean, and therefore the roll-off of $G$ tracks the loss of correlation with the linear field rather than an additional smearing of the acoustic feature, explaining why we have larger damping parameters in our priors in \cref{sec:priors}.
    \item The $r \to 0$ suppression acts on the field as a whole, and therefore it applies identically to the oscillatory and smooth components, and does not require separate handling of BAO and non-BAO damping.
    \item Since shot-noise-dominated modes are driven toward zero rather than propagated as white noise, the LiFT field carries no Poisson contribution to subtract or model separately in the covariance (shown explicitly in \cref{sec:cov,app:shot-noise}).
\end{enumerate} 

\section{Standard-reconstruction power spectrum fits}
\label{app:desi-pk-validation}

We verify that our standard reconstruction pipeline is faithful to the official analysis by comparing our post-reconstruction DR1 LRG fits against the Fourier-space measurements of appendix A of \cite{adame2025desi}. The comparison is end-to-end, as we run our own reconstruction on the DR1 catalogs (\cref{sec:standard-reconstruction}), measure our own multipoles (\cref{sec:pk}), and build our own analytic covariance (\cref{sec:cov}). \Cref{tab:desi_pk_validation} shows that the posterior means agree to better than $0.1\sigma$, and the posterior widths agree to within $2\%$. 

\begin{table}[t]
\centering
\caption{Our Fourier-space standard-reconstruction fits to the DESI DR1 LRG power spectra, compared with the corresponding DESI DR1 measurements. All uncertainties are statistical only.}
\label{tab:desi_pk_validation}
\setlength{\tabcolsep}{5pt}
\begin{tabular}{llccc}
\toprule
Bin & Analysis & $\alpha_{\rm iso}$ & $\alpha_{\rm AP}$ & $\chi^2/{\rm dof}$ \\
\midrule
\multirow{2}{*}{LRG1}
& This work & $0.9794\pm0.0116$ & $0.9235\pm0.0364$ & $84.6/87$ \\
& DESI DR1  & $0.9799\pm0.0118$ & $0.9231\pm0.0371$ & $85.9/87$ \\
\addlinespace
\multirow{2}{*}{LRG2}
& This work & $0.9625\pm0.0097$ & $1.0460\pm0.0354$ & $82.2/87$ \\
& DESI DR1  & $0.9624\pm0.0098$ & $1.0431\pm0.0356$ & $82.1/87$ \\
\addlinespace
\multirow{2}{*}{LRG3}
& This work & $1.0007\pm0.0083$ & $1.0061\pm0.0264$ & $102.4/87$ \\
& DESI DR1  & $1.0005\pm0.0081$ & $1.0057\pm0.0261$ & $100.8/87$ \\
\bottomrule
\end{tabular}
\end{table}

\section{Tied versus untied LiFT model response}
\label{app:untied}

In \cref{eq:LiFT-model}, we use a single response $B = G^2$ for both the smooth and wiggle components of the template. We verify that this simplification has no effect on the DESI results by refitting the DR1 multipoles with an untied model in which the smooth component receives its own damping, 
\begin{equation}
    B_{\rm nw}(k,\mu) = (b_0 + b_2\mu^2)^2 \exp[-k^2\mu^2\Sigma_{\rm nw}^2/2],
\end{equation}
with a Gaussian prior on $\Sigma_{\rm nw}$ of the same center and width as that on $\Sigma_\parallel$ (which we computed using the propagator in \cref{sec:priors}), and where $(k,\mu)$ is used for $B$ and $(k',\mu')$ is used for $C$, thus separating out the distortion as per the DESI convention. Ultimately, \cref{tab:untied} shows that the two models give the same result: the dilation parameters shift by at most $0.002$, well below their errors, and the errors themselves change by at most $0.001$. Moreover, the best-fit $\chi^2$ is unchanged to the quoted precision despite the additional parameter. 

\begin{table}[t]
\centering
\caption{Effect of freeing a separate damping $\Sigma_{\rm nw}$ for the smooth template component on the DESI DR1 LiFT fits. $\Delta\alpha$ and $\Delta\sigma$ are the untied minus tied posterior means and errors, and $\Delta\chi^2$ is the change in best-fit $\chi^2$ (the untied model has one fewer degree of freedom).}
\label{tab:untied}
\setlength{\tabcolsep}{5pt}
\begin{tabular}{lccccccc}
\toprule
Bin & $\Delta\alpha_{\rm iso}$ & $\Delta\sigma_{\rm iso}$ & $\Delta\alpha_{\rm AP}$ & $\Delta\sigma_{\rm AP}$ & $\Delta\chi^2$ & $\Sigma_{\rm nw}$ [$h^{-1}{\rm Mpc}$] \\
\midrule
LRG1 & $-0.0001$ & $0.0000$ & $+0.0001$ & $+0.0004$ & $0.0$ & $9.9\pm2.0$  \\
LRG2 & $+0.0000$ & $+0.0001$ & $+0.0024$ & $+0.0012$ & $0.0$ & $9.6\pm2.0$  \\
LRG3 & $+0.0004$ & $+0.0002$ & $-0.0007$ & $-0.0003$ & $0.0$ & $11.2\pm2.0$ \\
\bottomrule
\end{tabular}
\end{table}

\section{Shot noise in the LiFT field}
\label{app:shot-noise}

During the covariance computation in \cref{sec:cov}, we treat the LiFT field as a single continuous field rather than as a Poisson-sampled field for which there are shot-noise contributions, and therefore do not separately model shot noise contributions. We test this premise by measuring the excess kurtosis of the LiFT field as 
\begin{equation}
    g_2 \equiv \frac{\langle \delta^4 \rangle}{\langle \delta^2 \rangle^2} - 3.
    \label{eq:g2}
\end{equation}
For a Gaussian field with zero mean and variance $\sigma^2 = \langle\delta^2\rangle$, Wick's theorem gives
\begin{equation}
    \langle \delta^4 \rangle = 3\,\langle \delta^2 \rangle^2 = 3\sigma^4,
\end{equation}
so that $g_2 = 0$. We measure $g_2$ by evaluating the fields over all cells within the analysis mask at the native cell size of the grid, $5.2\,h^{-1}\mathrm{Mpc}$. Additionally, since a linear operation on a Gaussian field yields a Gaussian field, we also average over $2^3$ cells ($10.4\,h^{-1}\mathrm{Mpc}$), which should provide the same response. 

We compute $g_2$ on the DESI DR1 LRG reconstructed field from LiFT and on the meshed pre and post-standard-reconstruction galaxy fields. For reference, we also construct a Poisson null test by drawing an independent count $N$ in every cell with the same mean $\lambda$, thereby forming $\delta = N/\lambda - 1$ on the same mask as the data. We use this test instead of the theoretical value for $g_2$ of a homogenous Poisson process (which would give $1/\lambda \approx 40$ for $\lambda = 0.0247$) because in practice the observed field is very inhomogeneous over the mask: indeed, only $0.016\%$ of cells carry $90\%$ of $\langle\delta^4\rangle$, due to e.g. partially covered cells at the survey boundary and a strong drop-off in $n(z)$ in the LRG3 bin.

\begin{table}
    \centering
    \caption{Excess kurtosis $g_2$ (\cref{eq:g2}) of the DESI DR1 LRG fields. A Gaussian field has $g_2 = 0$ at both scales. The Poisson null is an independent Poisson count in each cell with the same mean count as the data, evaluated on the same mask.}
    \label{tab:kurtosis}
    \begin{tabular}{lcc}
        \toprule
        Field & $g_2$ ($5.2\,h^{-1}\mathrm{Mpc}$) & $g_2$ ($10.4\,h^{-1}\mathrm{Mpc}$) \\
        \midrule
        Galaxies & $1163 \pm 65$ & $157 \pm 13$ \\
        Standard reconstruction & $1155 \pm 50$ & $130 \pm 7$ \\
        Poisson null & $1192 \pm 32$ & $102 \pm 3$ \\
        LiFT & $0.58 \pm 0.06$ & $0.08 \pm 0.04$ \\
        \bottomrule
    \end{tabular}
\end{table}

Ultimately, we find that the galaxy field and standard reconstruction both sit roughly at the Poisson null test. By contrast, the LiFT field behaves almost the same as the Gaussian expectation, producing $g_2 = 0.58 \pm 0.06$ for one cell and $g_2 = 0.08 \pm 0.04$ for $2^3$ cells. These results establish that the LiFT field has massively reduced discreteness-induced contribution to its four-point function relative to standard reconstruction, and hence that the separate shot-noise terms of a Poisson-sampled covariance do not need to be applied to it. 

The test presented here only applies to the 
Poisson contribution to the covariance. 
Another contribution comes from the connected 
mode-coupling terms due to the non-linear evolution. These terms are dominated by the long wavelength mode couplings to the short wavelength modes. The super-sample variance 
term from modes outside the survey 
is expected to 
be small \cite{2014PhRvD..89h3519L}. The contribution from the 
modes within the survey is removed by LiFT
reconstruction of the Gaussian initial field, 
as demonstrated on periodic box simulations in \cite{bayer2026field}. These arguments, together with the overall coverage tests, provide confirmation that the LiFT posteriors are accurate. 

\section{Sensitivity to damping priors}
\label{app:prior-sensitivity}

We evaluate sensitivity to priors by refitting the LiFT DESI DR1 LRG power spectra with uniform rather than Gaussian priors. We use an equal-width interval as \cite{Paillas_2025}, centered on $[5,15]\,h^{-1}{\rm Mpc}$ for $(\Sigma_\parallel,\Sigma_\perp)$ in LiFT, and $b_0\in[0.01,5]$, $b_2\in[0,3]$. All other priors and fitting settings are held fixed. We include for comparison a refit of the standard reconstruction power spectrum using uniform priors of range $[0,10]\,h^{-1}{\rm Mpc}$ for $(\Sigma_\parallel,\Sigma_\perp,\Sigma_s)$ from \cite{Paillas_2025}, with its flat priors on $b_1$ and $\mathrm{d}\beta$ unchanged. \Cref{tab:prior-sensitivity} shows relatively small changes in the posterior means and uncertainties, with no material difference in response between standard reconstruction and LiFT.

\begin{table}[t]
\centering
\caption{Sensitivity to priors in our Fourier-space fits of the DESI DR1 LRG data. $\Delta\alpha$ and $\Delta\sigma$ are the flat minus Gaussian posterior means and errors, respectively. Standard reconstruction uses uniform damping priors of $[0,10]\,h^{-1}{\rm Mpc}$ and its same baseline priors on $b$ and $\mathrm{d}\beta$. LiFT uses $[5,15]\,h^{-1}{\rm Mpc}$ priors on the damping parameters and broad uniform priors of $b_0\in[0.01,5]$, $b_2\in[0,3]$. All other priors remain unchanged.}
\label{tab:prior-sensitivity}
\setlength{\tabcolsep}{5pt}
\begin{tabular}{lccccc}
\toprule
Bin & Method & $\Delta\alpha_{\rm iso}$ & $\Delta\sigma_{\rm iso}$
& $\Delta\alpha_{\rm AP}$ & $\Delta\sigma_{\rm AP}$ \\
\midrule
\multirow{2}{*}{LRG1}
& Standard & $+0.00025$ & $+0.00023$ & $-0.00020$ & $-0.00108$ \\
& LiFT     & $-0.00006$ & $+0.00008$ & $-0.00171$ & $+0.00019$ \\
\addlinespace
\multirow{2}{*}{LRG2}
& Standard & $+0.00084$ & $+0.00021$ & $-0.00251$ & $+0.00024$ \\
& LiFT     & $+0.00122$ & $+0.00004$ & $+0.00242$ & $-0.00036$ \\
\addlinespace
\multirow{2}{*}{LRG3}
& Standard & $-0.00008$ & $+0.00016$ & $-0.00106$ & $-0.00015$ \\
& LiFT     & $+0.00014$ & $-0.00007$ & $-0.00176$ & $+0.00005$ \\
\bottomrule
\end{tabular}
\end{table}

\section{Extended evaluation results}
\label{app:error_calibration}

\begin{figure}[p]
\centering
\includegraphics[width=\textwidth]{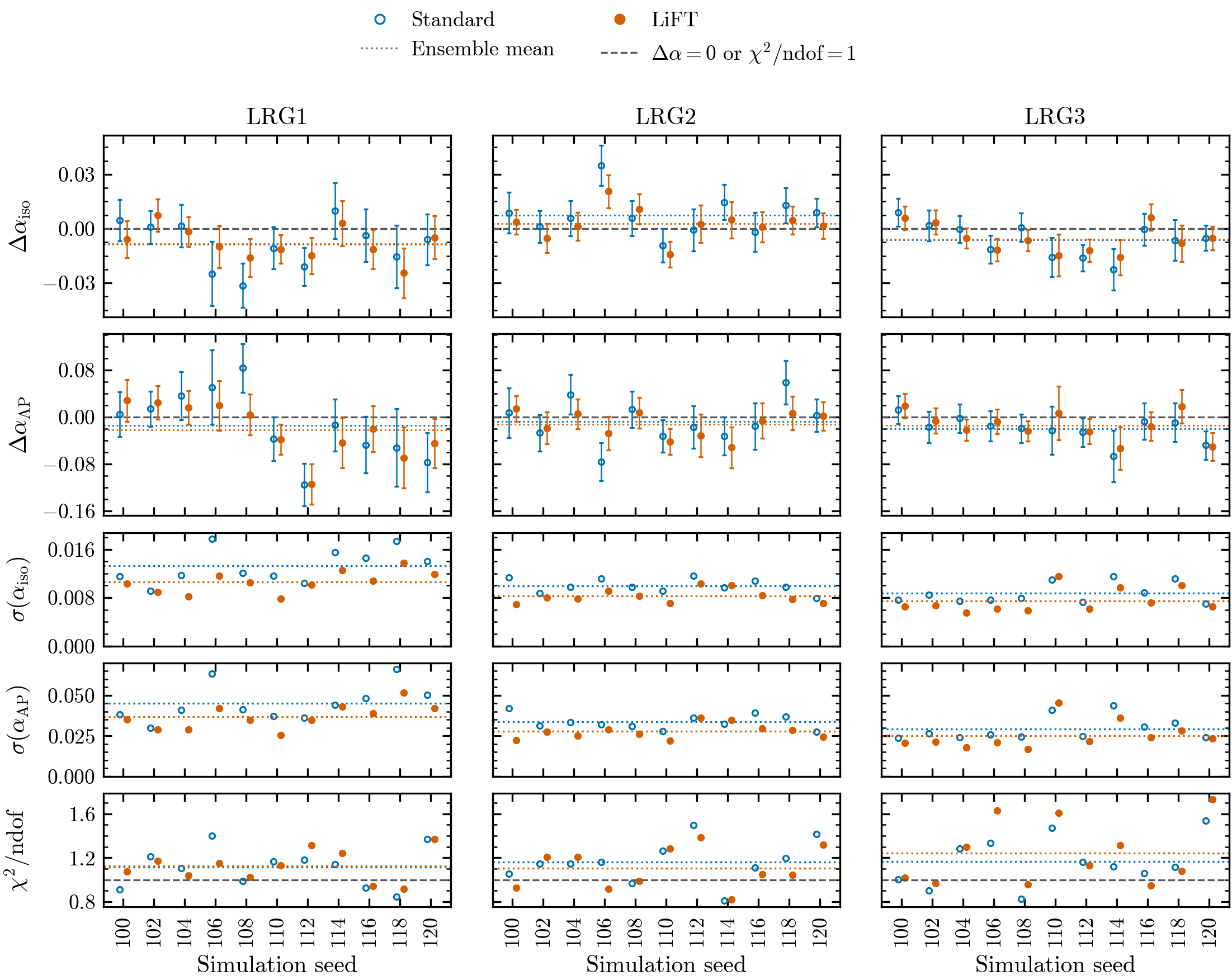}
\caption{BAO fits to each of the 11 HalfDome lightcones, comparing standard reconstruction (blue) and LiFT (orange). The first two rows, are the posterior means of $\alpha_{\rm iso}$ and $\alpha_{\rm AP}$ minus expectation, and the last rows shows $\chi^2/{\rm ndof}$.}
\label{fig:individual-halfdome}
\end{figure}

\begin{figure}[p]
\centering
\includegraphics[width=\textwidth]{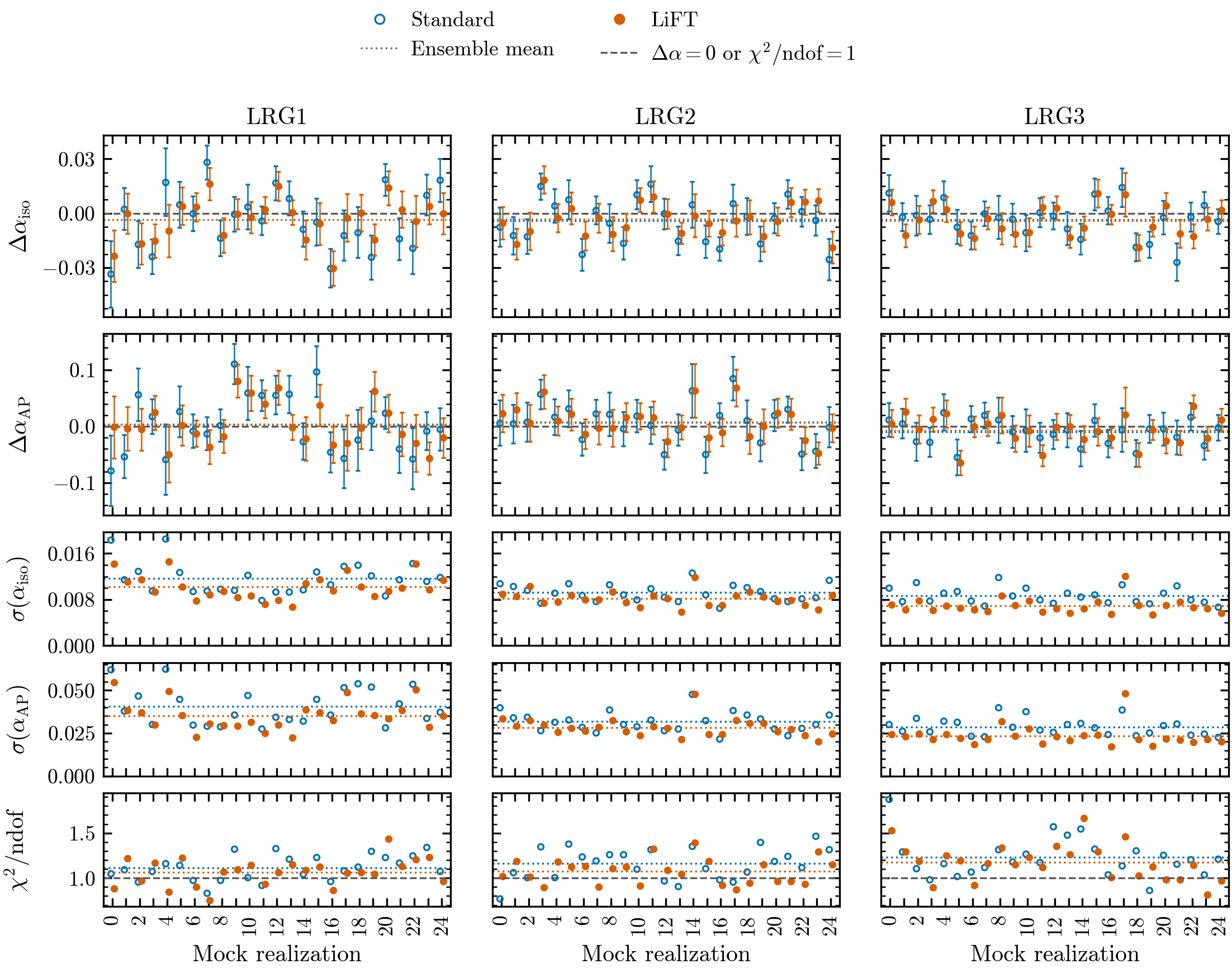}
\caption{Same as \cref{fig:individual-halfdome}, for the 25 AbacusSummit complete mocks.}
\label{fig:individual-complete}
\end{figure}

\begin{figure}[p]
\centering
\includegraphics[width=\textwidth]{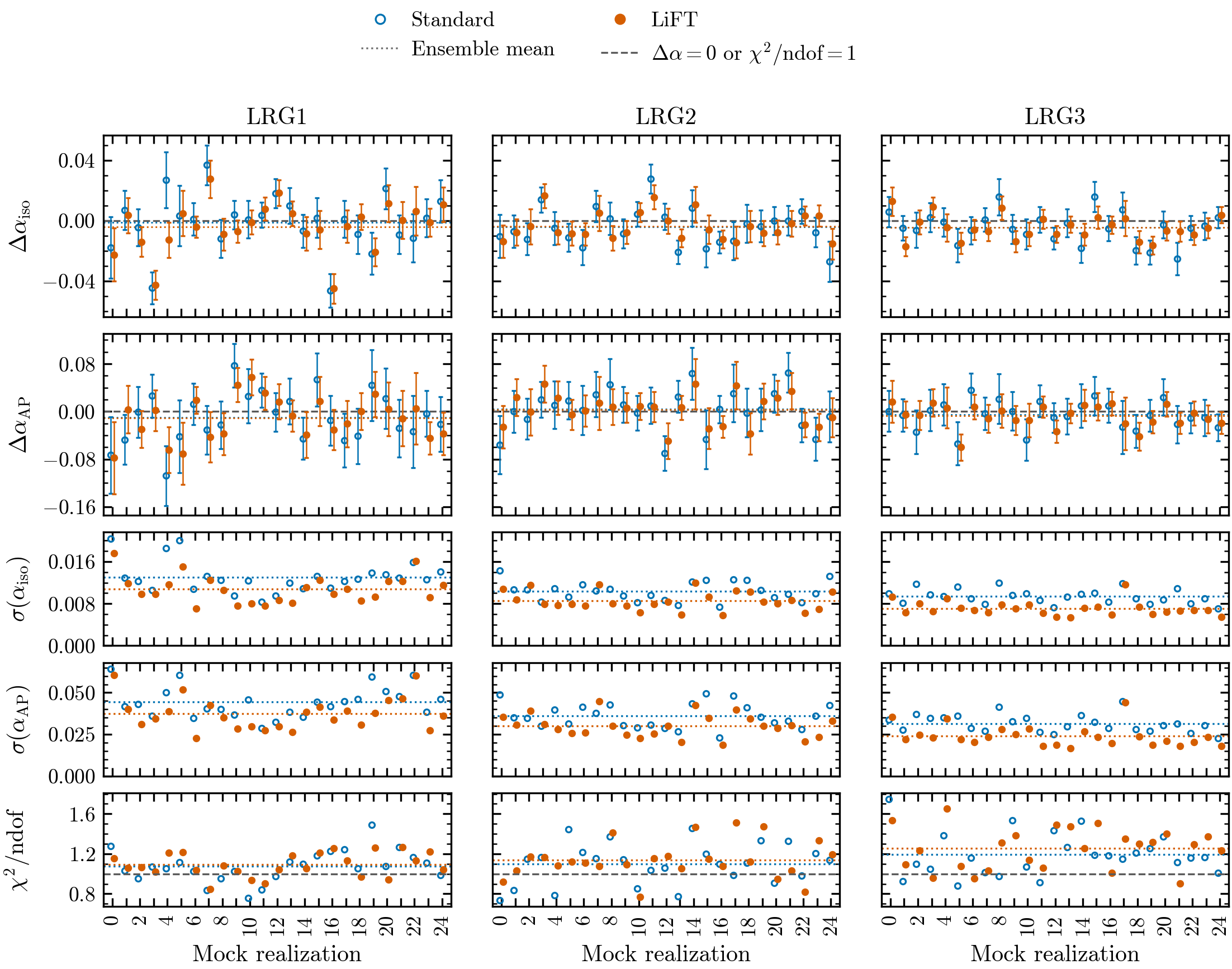}
\caption{Same as \cref{fig:individual-halfdome}, for the 25 AbacusSummit altMTL mocks. Mocks 4 and 17 sit at $3.5\sigma$ below the expected $\alpha_{\rm iso}$ in LRG1 for both methods and dominate the excess scatter of that bin in \cref{tab:ensemble-mse}.}
\label{fig:individual-altmtl}
\end{figure}

\begin{figure}[p]
\centering
\includegraphics[width=\textwidth]{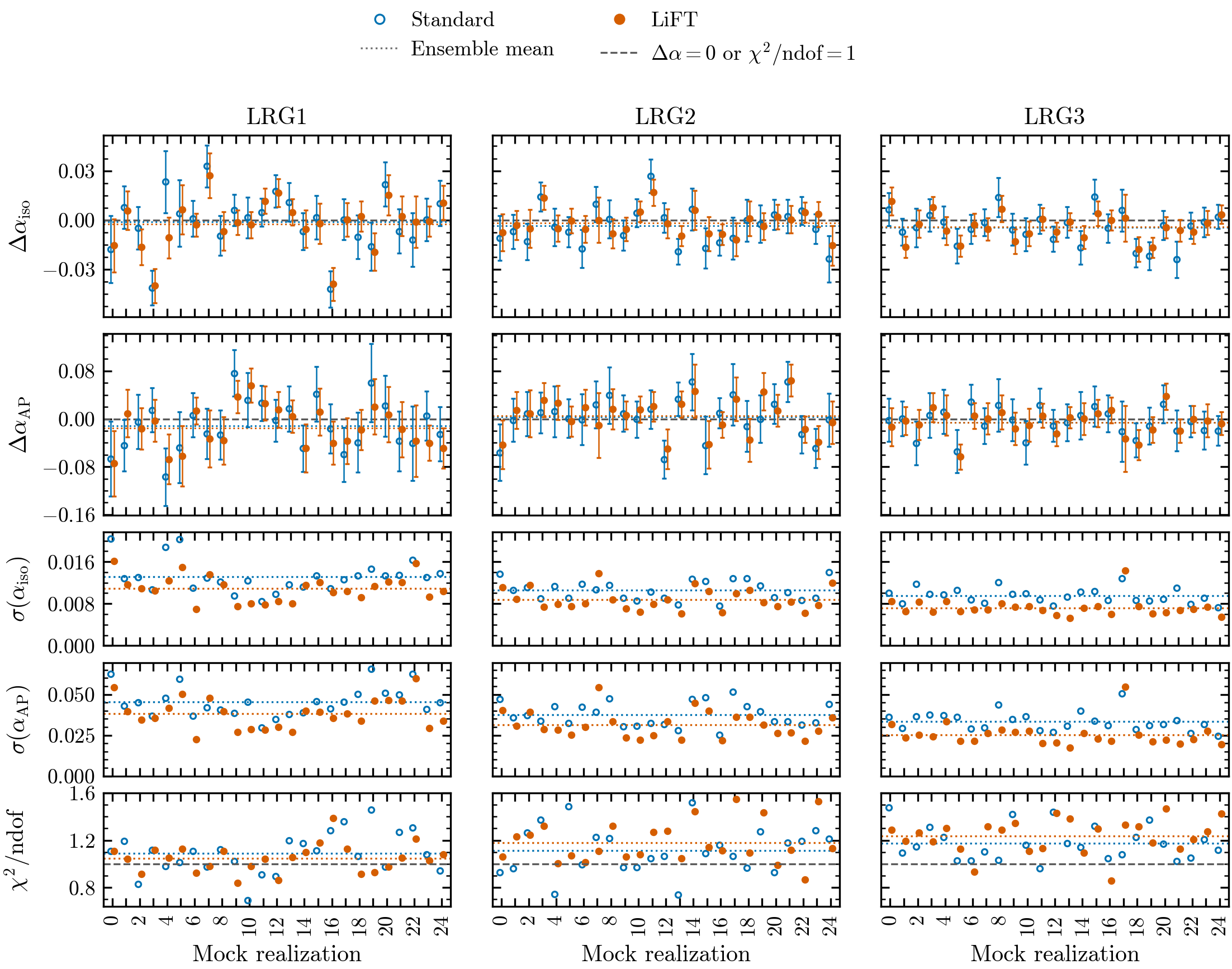}
\caption{Same as \cref{fig:individual-halfdome}, for the 25 AbacusSummit altMTL mocks analysed with the distorted distance-redshift mapping of \cref{sec:abacus-warped}; deviations are measured relative to the injected values of \cref{eq:alpha-expected}.}
\label{fig:individual-warped}
\end{figure}

In \Cref{tab:evaluation}, we provide the full numerical results that are generated during the held-out evaluation section. For each LRG bin, we report the mean $\alpha_{\rm iso}$ and $\alpha_{\rm AP}$ and the mean per-realization posterior width. We also show in \cref{fig:individual-halfdome,fig:individual-complete,fig:individual-altmtl,fig:individual-warped} the individual fits for each of the HalfDome or AbacusSummit simulations.

To compare empirical scatter with posterior widths, we combine the $11$ HalfDome and $25$ AbacusSummit altMTL mocks. We compute the standard deviation of the dilation parameters and compare them to the mean ensemble error in \Cref{tab:ensemble-mse}, where we define
\begin{equation}
\bar\alpha=\frac{1}{36}\sum_{i=1}^{36}\alpha_i,\qquad
s_\alpha=\left[\frac{1}{35}\sum_{i=1}^{36}(\alpha_i-\bar\alpha)^2\right]^{1/2},\qquad
\bar\sigma_\alpha=\frac{1}{36}\sum_{i=1}^{36}\sigma_{\alpha,i}.
\end{equation}
Averaging their ratio over the three bins and two parameters gives $1.06$ for LiFT and $1.08$ for standard reconstruction. We note that the agreement is not uniform across bins: in LRG1, the empirical scatter in $\alpha_{\rm iso}$ exceeds the mean posterior width by approximately $37\%$ for LiFT and $34\%$ for standard reconstruction. Much of this excess is driven by altMTL realizations 3 and 16, which recover unusually low $\alpha_{\rm iso}$ with both methods (\cref{fig:individual-altmtl}); omitting these lowers the LRG1 scatter-to-width ratio from $(1.37,1.34)$ to $(1.07,1.10)$ for LiFT and standard reconstruction respectively.

\begin{table}[p]
\centering
\caption{%
BAO constraints for the four evaluation tests described in \cref{sec:eval}. Entries in the mean-dilation columns are the ensemble mean $\pm$ the mean per-realization posterior width.} 
\label{tab:evaluation}
\small
\setlength{\tabcolsep}{4pt}
\begin{tabular}{llccccc}
\toprule
Bin & Method
& $\langle\alpha_{\rm iso}\rangle$
& $\langle\alpha_{\rm AP}\rangle$
& $\langle\chi^2/{\rm ndof}\rangle$
& $(\Delta\sigma_{\rm iso},\,\Delta\sigma_{\rm AP})$ & FoM \\
\midrule
\multicolumn{7}{l}{\textit{(a) Held-out HalfDome ($N=11$)}} \\
\addlinespace
\multirow{2}{*}{LRG1}
& Standard & $0.9912\pm0.0132$ & $0.9862\pm0.0451$ & 1.113 & --- & --- \\
& LiFT & $0.9918\pm0.0106$ & $0.9786\pm0.0369$ & 1.125 & $(20\%,\,18\%)$ & 1.5 \\
\addlinespace
\multirow{2}{*}{LRG2}
& Standard & $1.0074\pm0.0100$ & $0.9929\pm0.0336$ & 1.160 & --- & --- \\
& LiFT & $1.0029\pm0.0083$ & $0.9874\pm0.0278$ & 1.105 & $(17\%,\,17\%)$ & 1.5 \\
\addlinespace
\multirow{2}{*}{LRG3}
& Standard & $0.9939\pm0.0087$ & $0.9800\pm0.0293$ & 1.165 & --- & --- \\
& LiFT & $0.9942\pm0.0074$ & $0.9857\pm0.0252$ & 1.244 & $(15\%,\,14\%)$ & 1.4 \\
\midrule
\multicolumn{7}{l}{\textit{(b) AbacusSummit complete ($N=25$)}} \\
\addlinespace
\multirow{2}{*}{LRG1}
& Standard & $0.9965\pm0.0117$ & $1.0039\pm0.0407$ & 1.115 & --- & --- \\
& LiFT & $0.9963\pm0.0102$ & $1.0024\pm0.0353$ & 1.065 & $(13\%,\,13\%)$ & 1.3 \\
\addlinespace
\multirow{2}{*}{LRG2}
& Standard & $0.9960\pm0.0092$ & $1.0065\pm0.0317$ & 1.162 & --- & --- \\
& LiFT & $0.9967\pm0.0081$ & $1.0082\pm0.0282$ & 1.073 & $(11\%,\,11\%)$ & 1.3 \\
\addlinespace
\multirow{2}{*}{LRG3}
& Standard & $0.9966\pm0.0087$ & $0.9904\pm0.0287$ & 1.228 & --- & --- \\
& LiFT & $0.9960\pm0.0069$ & $0.9929\pm0.0233$ & 1.175 & $(21\%,\,19\%)$ & 1.6 \\
\midrule
\multicolumn{7}{l}{\textit{(c) AbacusSummit altMTL ($N=25$)}} \\
\addlinespace
\multirow{2}{*}{LRG1}
& Standard & $0.9986\pm0.0130$ & $0.9902\pm0.0443$ & 1.077 & --- & --- \\
& LiFT & $0.9959\pm0.0108$ & $0.9887\pm0.0374$ & 1.091 & $(17\%,\,16\%)$ & 1.4 \\
\addlinespace
\multirow{2}{*}{LRG2}
& Standard & $0.9957\pm0.0103$ & $1.0041\pm0.0360$ & 1.099 & --- & --- \\
& LiFT & $0.9963\pm0.0086$ & $1.0032\pm0.0300$ & 1.139 & $(17\%,\,17\%)$ & 1.4 \\
\addlinespace
\multirow{2}{*}{LRG3}
& Standard & $0.9954\pm0.0094$ & $0.9945\pm0.0315$ & 1.191 & --- & --- \\
& LiFT & $0.9953\pm0.0071$ & $0.9927\pm0.0238$ & 1.252 & $(25\%,\,24\%)$ & 1.8 \\
\midrule
\multicolumn{7}{l}{\textit{(d) AbacusSummit altMTL, AP distortion ($N=25$)}} \\
\addlinespace
\multirow{3}{*}{LRG1}
& Expected & $0.9615$ & $0.9706$ &  &  &  \\
& Standard & $0.9605\pm0.0132$ & $0.9593\pm0.0455$ & 1.088 & --- & --- \\
& LiFT & $0.9590\pm0.0109$ & $0.9554\pm0.0382$ & 1.045 & $(17\%,\,16\%)$ & 1.4 \\
\addlinespace
\multirow{3}{*}{LRG2}
& Expected & $0.9497$ & $0.9624$ &  &  &  \\
& Standard & $0.9462\pm0.0106$ & $0.9661\pm0.0376$ & 1.115 & --- & --- \\
& LiFT & $0.9478\pm0.0088$ & $0.9675\pm0.0313$ & 1.181 & $(17\%,\,17\%)$ & 1.4 \\
\addlinespace
\multirow{3}{*}{LRG3}
& Expected & $0.9396$ & $0.9561$ &  &  &  \\
& Standard & $0.9350\pm0.0095$ & $0.9504\pm0.0334$ & 1.175 & --- & --- \\
& LiFT & $0.9352\pm0.0072$ & $0.9499\pm0.0252$ & 1.235 & $(24\%,\,24\%)$ & 1.7 \\
\bottomrule
\end{tabular}
\end{table}

\begin{table}[p]
\centering
\caption{Empirical scatter and mean per-realization posterior widths for the combined $36$-mock HalfDome and AbacusSummit altMTL ensemble. The sample standard deviations are measured about the combined ensemble mean for each bin, parameter, and method. All mocks are retained.}
\label{tab:ensemble-mse}
\setlength{\tabcolsep}{5pt}
\begin{tabular}{llcccc}
\toprule
Bin & Method & $\mathrm{std}(\alpha_{\rm iso})$ & $\langle\sigma_{\rm iso}\rangle$ & $\mathrm{std}(\alpha_{\rm AP})$ & $\langle\sigma_{\rm AP}\rangle$ \\
\midrule
\multirow{2}{*}{LRG1}
& Standard & $0.0175$ & $0.0131$ & $0.0472$ & $0.0446$ \\
& LiFT & $0.0146$ & $0.0107$ & $0.0385$ & $0.0372$ \\
\addlinespace
\multirow{2}{*}{LRG2}
& Standard & $0.0130$ & $0.0102$ & $0.0348$ & $0.0353$ \\
& LiFT & $0.0093$ & $0.0085$ & $0.0260$ & $0.0293$ \\
\addlinespace
\multirow{2}{*}{LRG3}
& Standard & $0.0100$ & $0.0092$ & $0.0232$ & $0.0308$ \\
& LiFT & $0.0079$ & $0.0072$ & $0.0205$ & $0.0242$ \\
\bottomrule
\end{tabular}
\end{table}

\end{document}